\pdfoutput=1
\RequirePackage{fix-cm}
\documentclass[twocolumn]{svjour3}
\smartqed

\usepackage{amsmath,amsthm,amssymb,amsfonts,color,bm}
\usepackage{graphicx}
\usepackage{mathtools}
\usepackage{physics}
\usepackage{booktabs}
\usepackage{subfigure}
\usepackage{algorithmicx}
\usepackage{algorithm}
\usepackage{algpseudocode}
\algnewcommand\algorithmicinput{\textbf{Input:}}
\algnewcommand\Input{\item[\algorithmicinput]}
\algnewcommand\algorithmicoutput{\textbf{Output:}}
\algnewcommand\Output{\item[\algorithmicoutput]}
\usepackage[noadjust]{cite}
\definecolor{mypink1}{RGB}{142, 11, 11}

\newcommand{\cu}[1]{\mathbf{#1}}

\begin{document}
\sloppy 

\title{Kalman-based Spectro-Temporal ECG Analysis using Deep Convolutional Networks for Atrial Fibrillation Detection} \thanks{This work was supported by Business Finland}

\author{Zheng Zhao,         
        Simo S\"{a}rkk\"{a}, and
        Ali Bahrami Rad 
}

\institute{Department of Electrical Engineering and Automation\\
	Aalto University\\
	Rakentajanaukio 2c, Espoo\\
	02150, Finland
}

\date{Received: date / Accepted: date}

\maketitle

\begin{abstract}
In this article, we propose a novel ECG classification framework for atrial fibrillation (AF) detection using  spectro-temporal representation (i.e., time varying spectrum) and deep convolutional networks. In the first step we use a Bayesian spectro-temporal representation based on the estimation of time-varying coefficients of Fourier series using Kalman filter and smoother. Next, we derive an alternative model based on a stochastic oscillator differential equation to accelerate the estimation of the spectro-temporal representation in lengthy signals. Finally, after comparative evaluations of different convolutional architectures, we propose an efficient deep convolutional neural network to classify the 2D spectro-temporal ECG data.

The ECG spectro-temporal data are classified into four different classes: AF, non-AF normal rhythm (Normal), non-AF abnormal rhythm (Other), and noisy segments (Noisy). The performance of the proposed methods is evaluated and scored with the PhysioNet/Computing in Cardiology (CinC) 2017 dataset. The experimental results show that the proposed method achieves the overall F1 score of 80.2\%, which is in line with the state-of-the-art algorithms.


\end{abstract}

\section{Introduction}
\label{sec:intro}

Atrial fibrillation (AF) is the most common cardiac arrhythmia, and its prevalence is around 1--2\% worldwide~\cite{developed2010guidelines}. It is also estimated that by 2030 only in European Union 14--17 million patients suffer from AF~\cite{zoni2014epidemiology}. AF is associated with an increased risk of having stroke (5-fold), blood clots, heart failure, coronary artery disease, or death (2-fold; death rates are doubled by AF)~\cite{developed2010guidelines}. Therefore, developing automatic algorithms for early detection of AF is crucial.

During AF atrial muscle fibers have chaotic electrical activity which may emit impulses with 500 bpm rate to atrioventricular (AV) node, from which impulses pass randomly. This results to an irregular ventricular response which is one of the main characteristics of AF~\cite{thaler2017only}. In addition, AF has the following characteristics on electrocardiogram (ECG): 1) ``absolutely'' irregular RR intervals; 2) the absence of P waves; and 3) variable atrial cycle length (when visible). 

The analysis of ECG is the most common approach to AF detection, and during the past ten years, various algorithms have been developed for automatic AF detection~\cite{bruser2013automatic,yaghouby2010towards,asgari2015automatic,mohebbi2008detection,zabihi2017detection, ANNAVARAPU2016151, BABAEIZADEH2009522, GARCIA2016157, HAGIWARA201899}. Most of the existing algorithms follow a traditional pipeline of pre-processing, feature extraction, and classification. The recent deep learning (DL) techniques \cite{lecun2015deep} also provide a promising framework for end-to-end classification. In contrast to traditional approaches, one of the most significant advantages of using deep learning for classification is that hand-crafted features are no longer needed, because deep neural networks have the ability of learning the inherent features when provided with a sufficient training data \cite{goodfellow2016deep}. Whilst surprisingly, the applications of deep learning in AF have just begun in the past few years (see, e.g., \cite{rajpurkar2017cardiologist,shashikumar2017deep,xia2018detecting,pourbabaee2017deep, 8331569}).

For ECG signals, one can directly adopt 1D convolutional or recurrent network models for the classification task. However, transforming signals into spectral domain (spectro-temporal features) is a promising alternative approach knowing that the current state-of-the-art deep convolutional neural networks (CNNs) structures are typically designed for 2D images. Deep CNNs such as AlexNet \cite{NIPS2012_4824}, Inception-v4 \cite{AAAI1714806}, and DenseNet \cite{huang2017densely} have proved their superiority in image classification.

Within the previous studies, only a few have resorted to the use of time-varying spectrum for AF detection. The reasons might be the following. First, it is not easy to select hand-crafted features from 2D data using traditional classifiers. Second, the temporal features of spectrogram are usually hard to capture even in DL setting. Several studies \cite{xia2018detecting,zihlmann2017convolutional} have endeavoured DL for AF detection in spectral domain, but the use of traditional spectral estimation methods such as short-time Fourier transform (STFT) or continuous wavelet transform (CWT) may drop momentous information during the transformation, and produce less informative input data. Thus, to unravel these problems, it is beneficial to consider new spectro-temporal estimation methods that retain the temporal features better. 

The contributions of this paper are: 1) We propose two extended models for spectro-temporal estimation using Kalman filter and smoother. We then combine them with deep convolutional networks for AF detection. 2) We test and compare the performance of proposed approaches for spectro-temporal estimation on simulated data and AF detection with other popular estimation methods and different classifiers. 3) For AF detection, we evaluate the proposals using PhysioNet/CinC 2017 dataset~\cite{clifford2017af}, which is considered to be a challenging dataset that resembles practical applications, and our results are in line with the state-of-the-art. 

This paper is an extended version based of our previous conference paper ``Spectro-temporal ECG Analysis for Atrial Fibrillation Detection'' \cite{zhao2018spectro} presented at 2018 IEEE 28th International Workshop on Machine Learning for Signal Processing. In addition to the original contributions in the conference article, in this article, we use a new stochastic oscillator model and show that the spectro-temporal estimation can also be implemented with a steady state (stationary) Kalman filter and smoother, which leads to a significant reduction in time consumption without losing estimation accuracy. We demonstrate this in both simulated data and AF data classification. In addition to the experiments in the conference paper, where we only showed a few comparisons among estimation methods and classifiers, we expand them to a wide range of both standard and modern (e.g., Random Forests, CNNs, and DenseNet) classifiers for a better and more solid illustration of the classification performance.

The paper is structured as follows: In Section 2, we propose spectro-temporal methods for ECG signal analysis. In Section 3, we apply the proposed estimation method to AF detection using an averaging procedure. In Section 4, we compare and discuss experimental results both in simulated data and ECG dataset, followed by conclusion in Section 5.

\section{Spectro-Temporal Estimation Methods}
\label{sec:st-overall} 

Spectro-temporal signal analysis is an effective and powerful approach that is used in many fields ranging from biosignal analysis \cite{rad2017ecg} and audio processing \cite{rad2012phase} to weather forecasting \cite{ehrendorfer2012spectral}
and stock market prediction \cite{joseph2017daily}. In ECG analysis, the temporal evolution of spectral information can be captured in spectro-temporal data representation, which can convey important information about the underlying biological process of the heart.

In this section, we develop new methods for spectro-temporal estimation. We first introduce a Fourier series model based upon the Bayesian spectrum estimation method of Qi et al. \cite{qi2002bayesian}, and put Gaussian process priors on the Fourier coefficients. Then, by adopting the ideas presented in \cite{SARKKA20121517}, we convert the Fourier series into a more flexible stochastic oscillator model and use a fast stationary Kalman filter/smoother for its estimation. Finally, we demonstrate the estimation performance on simulated data.

\subsection{Kalman-based Fourier Series Model for Spectro-Temporal Estimation}
\label{sec:kalman}

Apart from traditional STFT and CWT methods, the spectro-temporal analysis can also be done by modeling the signal as a stochastic state-space model and resorting to the Bayesian procedure (i.e., Kalman filter and smoother) for its estimation \cite{sarkka2013bayesian,qi2002bayesian}. The key advantages of this kind of approaches over other spectro-temporal methods are that we can apply them to both evenly and unevenly sampled signals \cite{qi2002bayesian} and they require no stationarity guarantees nor windowing. Furthermore, as we show here, they can also be combined with state-space methods for Gaussian processes \cite{Hartikainen+Sarkka:2010,Sarkka+Solin+Hartikainen:2013}. 

Recall that any periodic signal with fundamental frequency $f_0$ can be expanded into a Fourier series
\begin{equation}
z(t) = a_0 + \sum_{j=1}^{M} \left[ a_{j} \cos(2\pi \, j \, f_0 t) + b_{j} \sin(2\pi  \, j \, f_0 \, t) \right],
\label{equ:fourier_series}
\end{equation}
where the exact representation is obtained with $M \to \infty$, but for sampled (and thus bandlimited) signals it is sufficient to consider finite series. This stationary model is the underlying model in the STFT approach. STFT applies a window to each signal segment and finds a least squares fit (via discrete Fourier transform) to the coefficients $\{ a_j, b_j : j=1,\ldots,M \}$.

In our approach, we start by assuming that the coefficients depend on time, and we put Gaussian process priors on them:
\begin{equation}
\begin{split}
a_j(t) &\sim \mathcal{GP}(0,k^a_j(t,t')), \\
b_j(t) &\sim \mathcal{GP}(0,k^b_j(t,t')).
\end{split}
\end{equation}
As shown in \cite{Hartikainen+Sarkka:2010,Sarkka+Solin+Hartikainen:2013}, provided that the covariance functions are stationary, we can express the Gaussian processes as solutions to linear stochastic differential equations (SDEs). We choose the covariance functions to have the form
\begin{equation}
\begin{split}
k^a_j(t,t') &= (s^a_j)^2 \, \exp( -\lambda^a_j \, | t - t' | ), \\
k^a_j(t,t') &= (s^b_j)^2 \, \exp( -\lambda^b_j \, | t - t' | ),
\end{split}
\end{equation}
where $s^a_j,s^b_j > 0$ are scale parameters and $\lambda^a_j,\lambda^b_j > 0$ are the inverses of the time constants (length scales) of the processes. 

The state-space representations (which are scalar in this case) are then given as
\begin{equation}
\begin{split}
da_j &= -\lambda^a_j \, a_j \, dt + dW^a_j, \\
db_j &= -\lambda^b_j \, b_j \, dt + dW^b_j, \\
\end{split}
\end{equation}
where $W^a_j,W^b_j$ are Brownian motions with suitable diffusion coefficients $q^a_j,q^b_j$. We can also solve the equations at discrete time steps (see, e.g., \cite{Grewal+Andrews:2001}) as
\begin{equation}
\begin{split}
a_j(t_k) &= \psi^a_{jk} \, a_j(t_{k-1}) + w^a_{jk},  \quad w^a_{jk} \sim \mathcal{N}(0,\Sigma^a_{jk}),\\
b_j(t_k) &= \psi^b_{jk} \, b_j(t_{k-1}) + w^b_{jk},  \quad w^b_{jk} \sim \mathcal{N}(0,\Sigma^b_{jk}),\\
\end{split}
\label{eq:discdyn}
\end{equation}
where
\begin{equation}
\begin{split}
  \psi^a_{jk} &= \exp(-\lambda^a_j \, (t_k - t_{k-1})), \\
  \psi^b_{jk} &= \exp(-\lambda^b_j \, (t_k - t_{k-1})), \\
  \Sigma^a_{jk} &= q^a_j \, (1 - \exp(-2 \lambda^a_j \, (t_k - t_{k-1}))), \\
  \Sigma^b_{jk} &= q^b_j \, (1 - \exp(-2 \lambda^b_j \, (t_k - t_{k-1}))).
\end{split}
\end{equation}
Let us now assume that we obtain noisy measurements of the Fourier series \eqref{equ:fourier_series} at times $t_1,t_2,\ldots$. What we can now do is to define a state vector $\mathbf{x} = [a_{0}, a_{1}, ..., a_{M}, b_{1}, b_{2}, \ldots, b_{M}]^\top$ which stacks all the coefficients $a_j$ and $b_j$. In this way, we can write  $\mathbf{H}_k = [1, \cos(2\pi f_0 t_k), \ldots, \cos(2\pi M \, f_0 \, t_k), \sin(2\pi f_0 t_k), \ldots, \\\sin(2\pi Mf_0 t_k)]$, which leads to 
\begin{equation}
\begin{split}
z(t_k) &=  a_0 + \sum_{j=1}^{M} \left[ a_{j} \cos(2\pi \, j \, f_0 t_k) + b_{j} \sin(2\pi  \, j \, f_0 \, t_k) \right]\\
&=\mathbf{H}_k \, \mathbf{x}_k.
\end{split}
\raisetag{2\baselineskip}
\label{eq:Hx}
\end{equation}
We can also rewrite the dynamic model \eqref{eq:discdyn} as
\begin{equation}
\mathbf{x}_k = \bm{\mathrm{\Psi}}_{k} \, \mathbf{x}_{k-1} + \mathbf{q}_{k},
\label{eq:dynmodel}
\end{equation}
where $\bm{\mathrm{\Psi}}_k$ contains the terms $\psi^a_{jk}$ and $\psi^b_{jk}$ on the diagonal and $ \mathbf{q}_{k} \sim \mathcal{N}(\mathbf{0},\bm{\mathrm{\Sigma}}_k)$ where $\bm{\mathrm{\Sigma}}_k$ contains the terms $\Sigma^a_{jk}$ and $\Sigma^b_{jk}$ on the diagonal.

If we assume that we actually measure \eqref{eq:Hx} with additive Gaussian measurement noise $r_k \sim \mathcal{N}(0,R)$, then we can express the measurement model as
\begin{equation}
\begin{split}
y_k &= \mathbf{H}_k \, \mathbf{x}_k + r_k.
\end{split}
\label{eq:measmodel}
\end{equation}
Equations \eqref{eq:dynmodel} and \eqref{eq:measmodel} define a linear state-space model where we can perform exact Bayesian estimation using Kalman filter and smoother \cite{sarkka2013bayesian}. In the original paper \cite{qi2002bayesian}, the state vectors $\mathbf{x}_1, ..., \mathbf{x}_N$ are assumed to perform random walk, but here the key insight is to use a more general Gaussian process which introduces a finite time constant to the problem. Although here we have chosen to use quite simple Gaussian process model for this purpose, it would also be possible to use more general Gaussian process priors for the coefficients such as state-space representations of Mat\'ern or squared exponential covariance functions \cite{Hartikainen+Sarkka:2010,Sarkka+Solin+Hartikainen:2013}. 

The Kalman filter for this problem then consists of the following forward recursion (for $k=1,\ldots,N$):

\begin{equation}
\begin{split}
\mathbf{m}^-_{k} &= \bm{\mathrm{\Psi}}_{k} \, \mathbf{m}_{k-1}, \\
\mathbf{P}^-_{k} &= \bm{\mathrm{\Psi}}_{k} \, \mathbf{P}_{k-1} \, \bm{\mathrm{\Psi}}_{k}^\top
+ \bm{\mathrm{\Sigma}}_{k}, \\
S_{k} &= \mathbf{H}_{k} \, \mathbf{P}^-_{k} \, \mathbf{H}_{k}^\top + R, \\
\mathbf{K}_{k} &= \mathbf{P}^-_{k} \, \mathbf{H}^\top_{k} / S_{k}, \\
\mathbf{m}_{k} &= \mathbf{m}^-_{k} + \mathbf{K}_{k} \, \left( y_k - \mathbf{H}_{k} \, \mathbf{m}^-_{k} \right), \\
\mathbf{P}_{k} &= \mathbf{P}^-_{k} - \mathbf{K}_{k} \, S_{k} \, \mathbf{K}^\top_{k}, \\
\end{split}
\label{eq:dkf_update}
\end{equation}
and the RTS smoother the following backward recursion (for $k=N-1,\ldots,1$):
\begin{equation}
\begin{split}
\mathbf{G}_{k} &= \mathbf{P}_k \, \bm{\mathrm{\Psi}}_{k+1}^\top \, [\mathbf{P}^-_{k+1}]^{-1}, \\
\mathbf{m}^\mathrm{s}_k &= \mathbf{m}_k
+ \mathbf{G}_k \, [\mathbf{m}^\mathrm{s}_{k+1} - \mathbf{m}^-_{k+1}], \\
\mathbf{P}^\mathrm{s}_k &= \mathbf{P}_k
+ \mathbf{G}_k \, [\mathbf{P}^\mathrm{s}_{k+1} - \mathbf{P}^-_{k+1}] \, \mathbf{G}^\top_k.
\end{split}
\label{eq:kfs}
\end{equation}
The final posterior distributions are then given as:
\begin{equation}
p(\mathbf{x}_{k} \mid y_{1:N}) =
\mathcal{N}(\mathbf{x}_{k} \mid \mathbf{m}^\mathrm{s}_{k},\mathbf{P}^\mathrm{s}_{k}),
\quad k = 1,\ldots,N.
\label{eq:kfs_p}
\end{equation}
%
The magnitude of the sinusoidal with frequency $f_j = j \, f_0 $ at time step $k$ can then be computed by extracting the elements corresponding to $\hat{a}_{j}(t_k)$ and $\hat{b}_{j}(t_k)$ from the mean vector $\mathbf{m}^\mathrm{s}_{k}$:
\begin{equation}
[\mathbf{S}]_{j,k} = \sqrt{\hat{a}_{j}^2(t_k) + \hat{b}_{j}^2(t_k)}.
\label{eq:power_density}
\end{equation}
From now on, matrix $\mathbf{S}$ is called spectro-temporal data matrix.

\subsection{Oscillator Model for Spectro-Temporal Estimation}
\label{sec:osc}

In practice, the computational cost of Kalman filter and smoother can be extensive when the length of the signal is very long. However, instead of the Fourier series state space model in previous section, one can also derive an alternative representation using stochastic oscillator differential equations. In this way, the dynamic and measurement models become linear time-invariant (LTI) so that we can leverage a stationary Kalman filter to reduce the time consumption. This kind of stochastic oscillator models were also considered in \cite{SARKKA20121517} and the link to period Gaussian process models was investigated in \cite{solin14}. 

A single quasi-period stochastic oscillator can be described with the following stochastic differential equation model \cite{solin14}:
\begin{equation}
\begin{split}
d\cu{x}^j&= \begin{bmatrix}
-\lambda_j & -2\pi f_j\\
2\pi f_j & -\lambda_j
\end{bmatrix}\,\cu{x}^j \, dt + \begin{bmatrix}
1 & 0\\0 & 1
\end{bmatrix} \, d\cu{W}_j, \\
&= \cu{F}_j \,\cu{x}^j \, dt + \cu{L}\,d\cu{W}_j.
\label{equ:osc-sde}
\end{split}
\end{equation}
where $\cu{x}^j=\begin{bmatrix}
a_j & b_j
\end{bmatrix}^\top$ and the Brownian motion $\cu{W}_j = \begin{bmatrix}
W^a_j & W^b_j
\end{bmatrix}^\top$ has a suitably chosen diffusion matrix $\bm{\zeta}^j = q_j \, \cu{I}$ \cite{solin14}. By solving the SDE in discrete time steps, we have 
\begin{equation}
\cu{x}^j_k = \cu{A}^j \, \cu{x}^j_{k-1} + \cu{q}^j, \quad \cu{q}^j\sim \mathcal{N}(\cu{0}, \cu{Q}^j),
\label{equ:osc-dis}
\end{equation}
where $\cu{A}^j$ and $\cu{Q}^j$ are given by:
\begin{equation}
\begin{split}
\cu{A}^j &= \exp(\cu{F}_j \, \Delta t), \\
\cu{Q}^j &= \int_0^{\Delta t} \exp(\cu{F}_j\, (\Delta t-s)) \, \cu{L}\, \bm{\zeta}^j\, \cu{L}^\top\\ &\qquad \quad \times \exp(\cu{F}_j\, (\Delta t-s))^\top \, ds,
\label{equ:osc-solve}
\end{split}
\end{equation}
where $\Delta t = t_k - t_{k-1}$.

A general quasi-periodic signal can be modeled using a superposition of stochastic oscillators of the above form \cite{solin14}. If we construct $\cu{x}_k = \begin{bmatrix} (\cu{x}^0_k)^\top & (\cu{x}^1_k)^\top & \cdots & (\cu{x}^M_k)^\top \end{bmatrix}^\top$, then the resulting time-invariant model can be written as:
\begin{align}
\cu{x}_k &=  \cu{A}\, \cu{x}_{k-1} + \cu{q}_k, \quad\cu{q}_k\sim\mathcal{N}(\cu{0},\cu{Q}),  \nonumber \\
y_k &= \cu{H} \, \cu{x}_k + r_k, \quad \quad \ r_k\sim\mathcal{N}(0, R).
\label{equ:osc-model}
\end{align}
where $\cu{A}$, $\cu{Q}$ and $\cu{H}$ are defined as:
\begin{align}
\cu{A} &= \begin{bmatrix}
1 & & &\\
 & \cu{A}^1 & &\\
 & & \ddots &\\
 & & & \cu{A}^M
\end{bmatrix}, \quad
\cu{Q} = \begin{bmatrix}
q_b \, \Delta t & & &\\
 & \cu{Q}^1 & & \\
 & & \ddots & \\
 & & & \cu{Q}^M
\end{bmatrix}, \\
\cu{H} &= \begin{bmatrix}
  1& \cu{H}^1& \cdots & \cu{H}^M
  \end{bmatrix}
  = \begin{bmatrix}
  1 & 1&  0&  1&  0& \cdots & 1&  0
  \end{bmatrix}.
\label{equ:osc-a-q-h}
\end{align}  
In this model, the first component of the state is a slowly drifting Brownian motion with diffusion coefficient $q_b$ modeling the possible non-zero mean of the signal.

The estimation problem can be solved with a Kalman filter and smoother. However, because the model is LTI, the Kalman filter is known to converge to a steady-state Kalman filter \cite{Kailath:233814}. The steady-state Kalman filter can be obtained by solving the following discrete algebraic Riccati equation (DARE) for the limit covariance $\cu{P}^-_k \to \cu{P}^-_\infty$:
\begin{equation}
\begin{split}
 \cu{P}^-_\infty
 &= \cu{A} \, \cu{P}_\infty^- \, \cu{A}^\top + \cu{Q}\\
 &- \cu{A} \, \cu{P}_\infty^- \, \cu{H}^\top  \, (\mathbf{H} \, \mathbf{P}^-_\infty \, \mathbf{H}^\top + R)^{-1}  \, \cu{H}  \, \cu{P}_\infty^-  \, \cu{A}^\top.
\end{split}
\label{equ:riccati-rc}
\end{equation}
A positive-semi-definite solution to the equation is known to exists provided that the pair $\left[ \cu{A}, \cu{H}\right] $ is detectable \cite{Kailath:233814}. 

Thus we can obtain $\cu{P}^-_\infty$ by solving DARE in  \eqref{equ:riccati-rc}, and the stationary Kalman filter for the forward mean propagation is:
\begin{equation}
\begin{split}
\cu{m}_k &= \cu{A} \, \cu{m}_{k-1} + \cu{K} \, (y_k - \cu{H} \, \cu{A} \, \cu{m}_{k-1}),
\end{split}
\label{equ:osc-forward}
\end{equation}
where the stationary gain is
\begin{equation}
\begin{split}
\cu{K} &= \cu{P}^-_\infty \, \cu{H}^\top  \, (\cu{H} \, \cu{P}^-_\infty \, \cu{H}^\top + R)^{-1}.\\
\end{split}
\end{equation}
The corresponding smoother then turns out to converge to its steady state as well, and the backward propagation for the resulting steady-state smoother is:
\begin{equation}
\begin{split}
\cu{m}^s_k &= \cu{m}_k + \cu{G} \, (\cu{m}^s_{k+1} - \cu{A} \, \cu{m}_{k}).
\end{split}
\end{equation}
where the gain is computed as
\begin{equation}
\begin{split}
\cu{G} &= \cu{P}_\infty \, \cu{A}^\top  \, [\cu{P}^-_\infty]^{-1}, \\
\cu{P}_\infty &= \cu{P}^-_\infty - \cu{P}^-_\infty \, \cu{H}^\top  \, (\cu{H} \, \cu{P}^-_\infty \, \cu{H}^\top + R)^{-1}  \, 
\cu{H} \, \cu{P}^-_\infty.\\
\end{split}
\end{equation}

In this way, the calculation of the filter and covariances at every time step is not needed, which reduces the computational cost significantly. The disadvantage is that we need to solve the DARE in order to construct the stationary filter and smoother, which also adds to the computational cost.

After computing the estimates $\cu{m}^s_k$ for each time step, we can extract the estimates of $\hat{a}_{j}(t_k)$ and $\hat{b}_{j}(t_k)$ and use \eqref{eq:power_density} to compute the spectro-temporal data matrix. 

\subsection{Estimation Trials on Simulated Data}

A quantitative evaluation of the proposed spectro-temporal methods for ECG classification is discussed in Sections \ref{results1} and \ref{MethodComparison}. However, in this section we visually inspect the proposed spectro-temporal representations on the simulated data and compare them with other standard time-frequency approaches such as STFT, CWT, and BurgAR.   
 
To avoid confusion in terminology, from now on, we refer the proposals in Section \ref{sec:kalman} and \ref{sec:osc} as FourierKS and OscKS, respectively.

\begin{figure}[t!]
	\centering
	\includegraphics[width=0.8\linewidth]{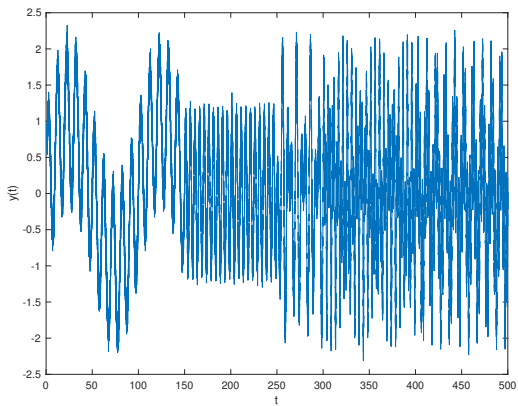}
	\caption{Simulated sinusoidal data.}
	\label{fig:simu-yt}
\end{figure}

\begin{figure}[htb!]
	\centering
	\subfigure[Kalman $\lambda=0.01$, $M_0=0, M_{50}=0.5$]{\label{fig:simu-kf}\includegraphics[width=0.48\linewidth]{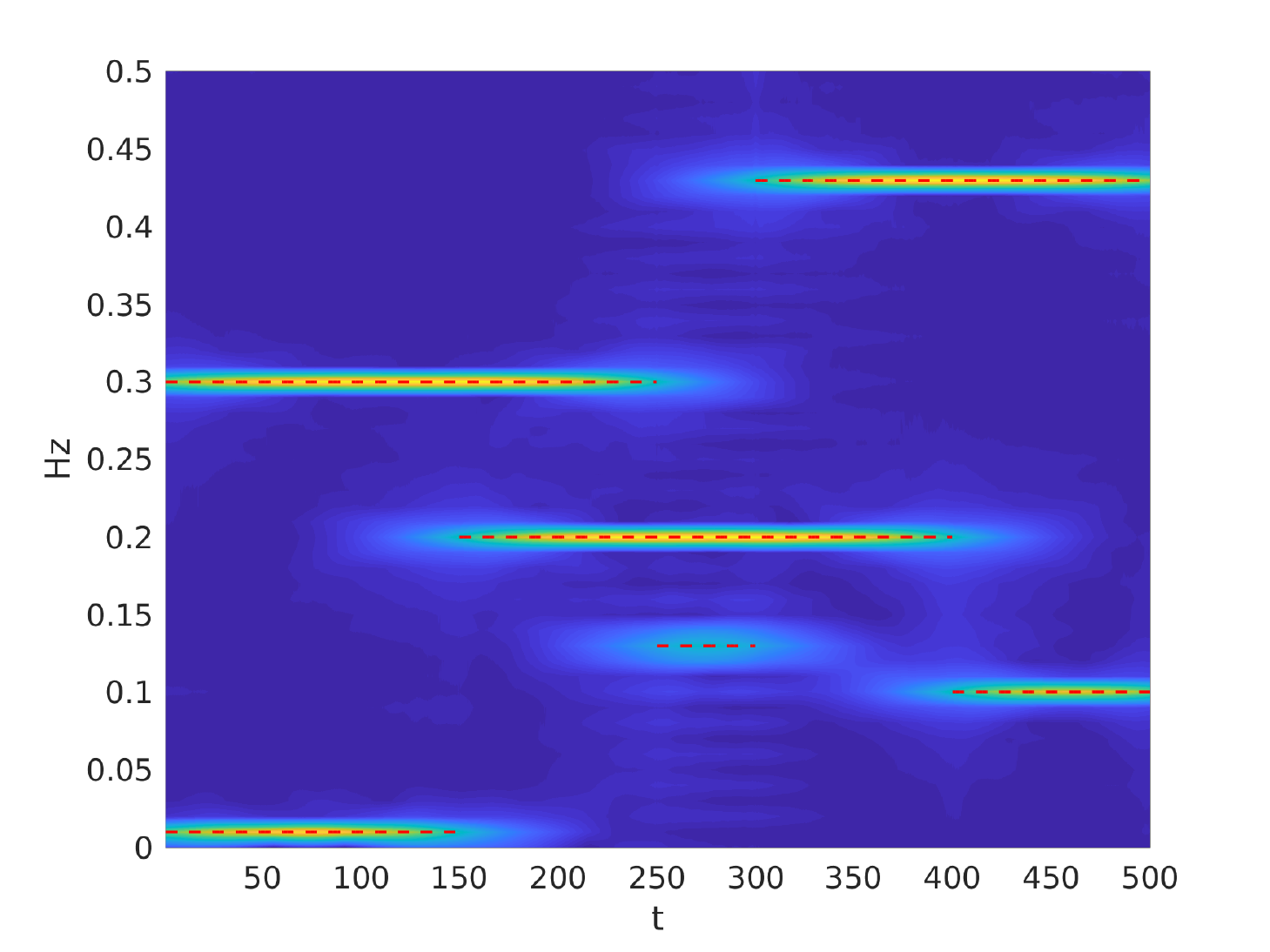}}
	\subfigure[OSC $\lambda=0.01$, $M_0=0, M_{50}=0.5$]{\label{fig:simu-osc}\includegraphics[width=0.46\linewidth]{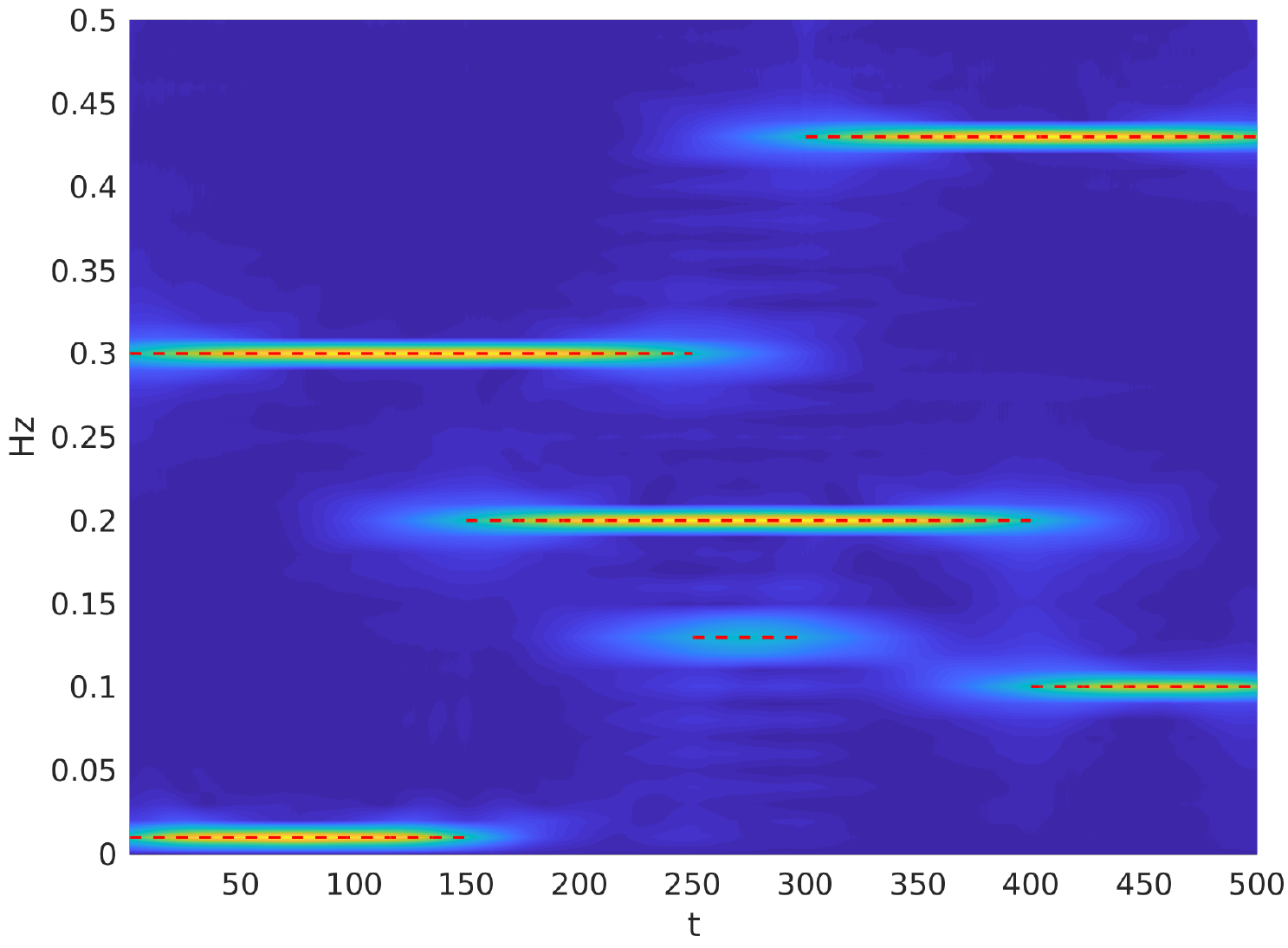}}
	\subfigure[CWT Morse wavelet, time bandwidth 60]{\label{fig:simu-cwt}\includegraphics[width=0.45\linewidth]{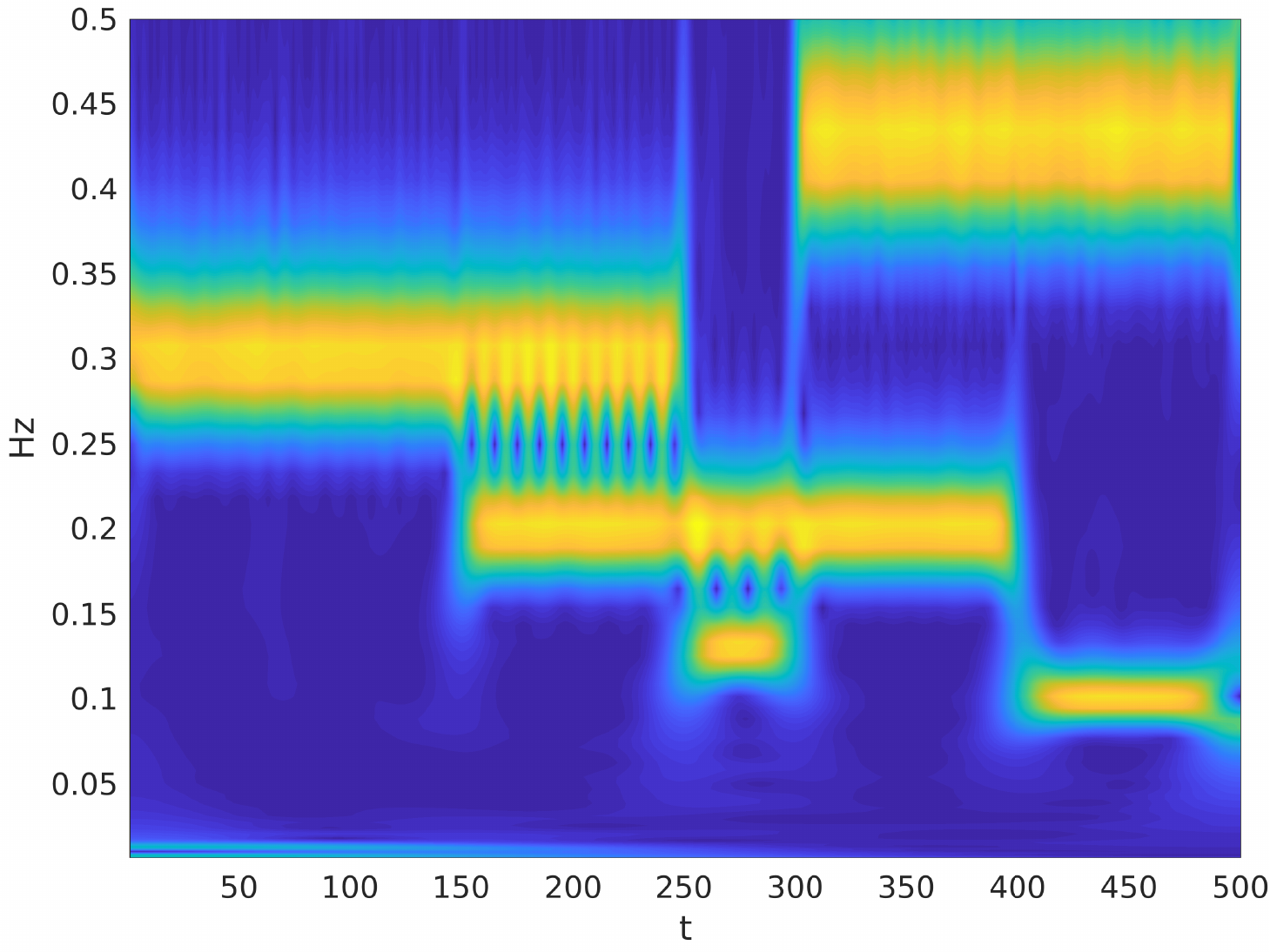}}
	\subfigure[STFT, Hann, window 350, overlap 340 ]{\label{fig:simu-stft}\includegraphics[width=0.46\linewidth]{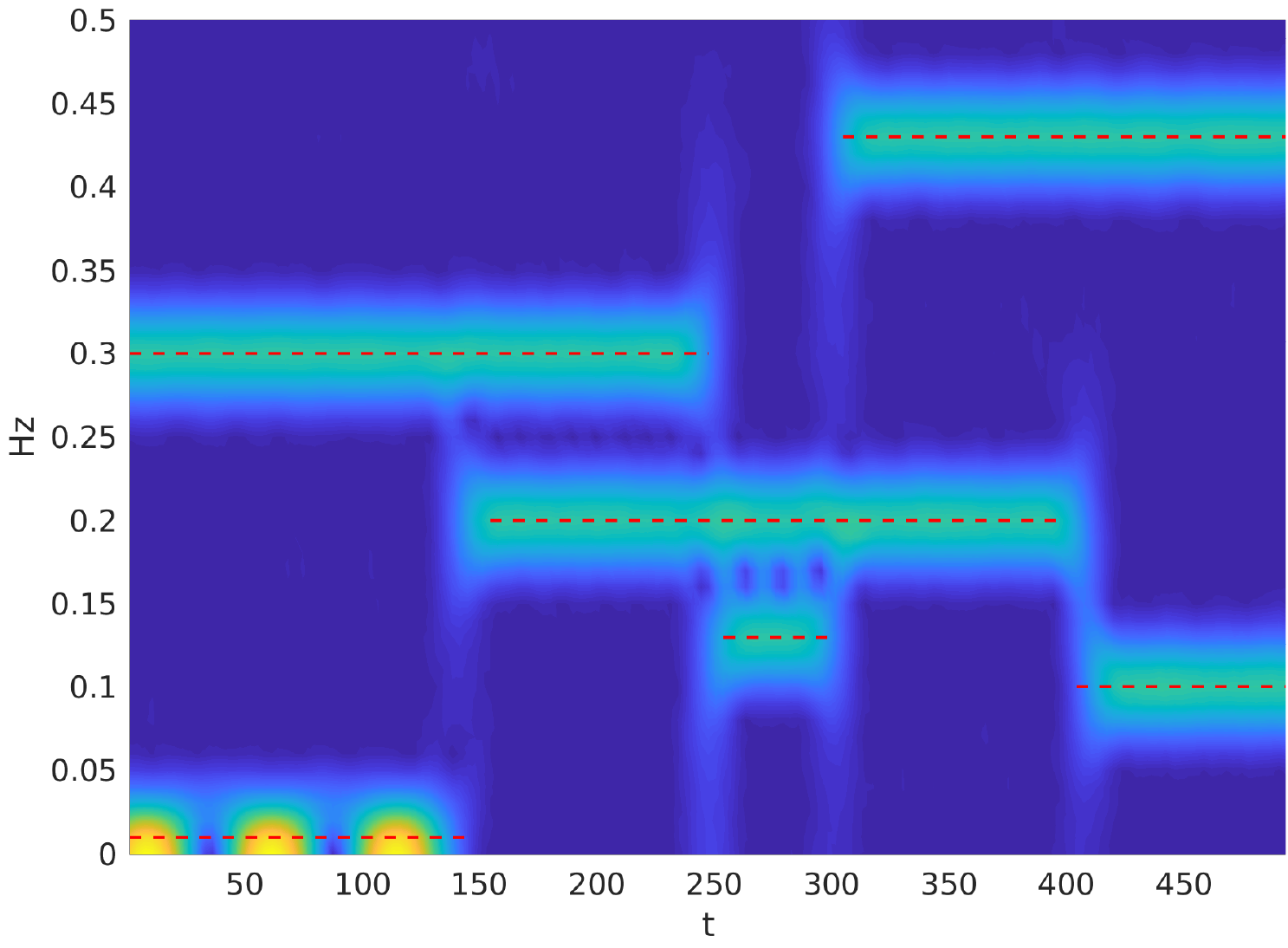}}
	\subfigure[BurgAR, Hann window 350, overlap 340]{\label{fig:simu-stft}\includegraphics[width=0.46\linewidth]{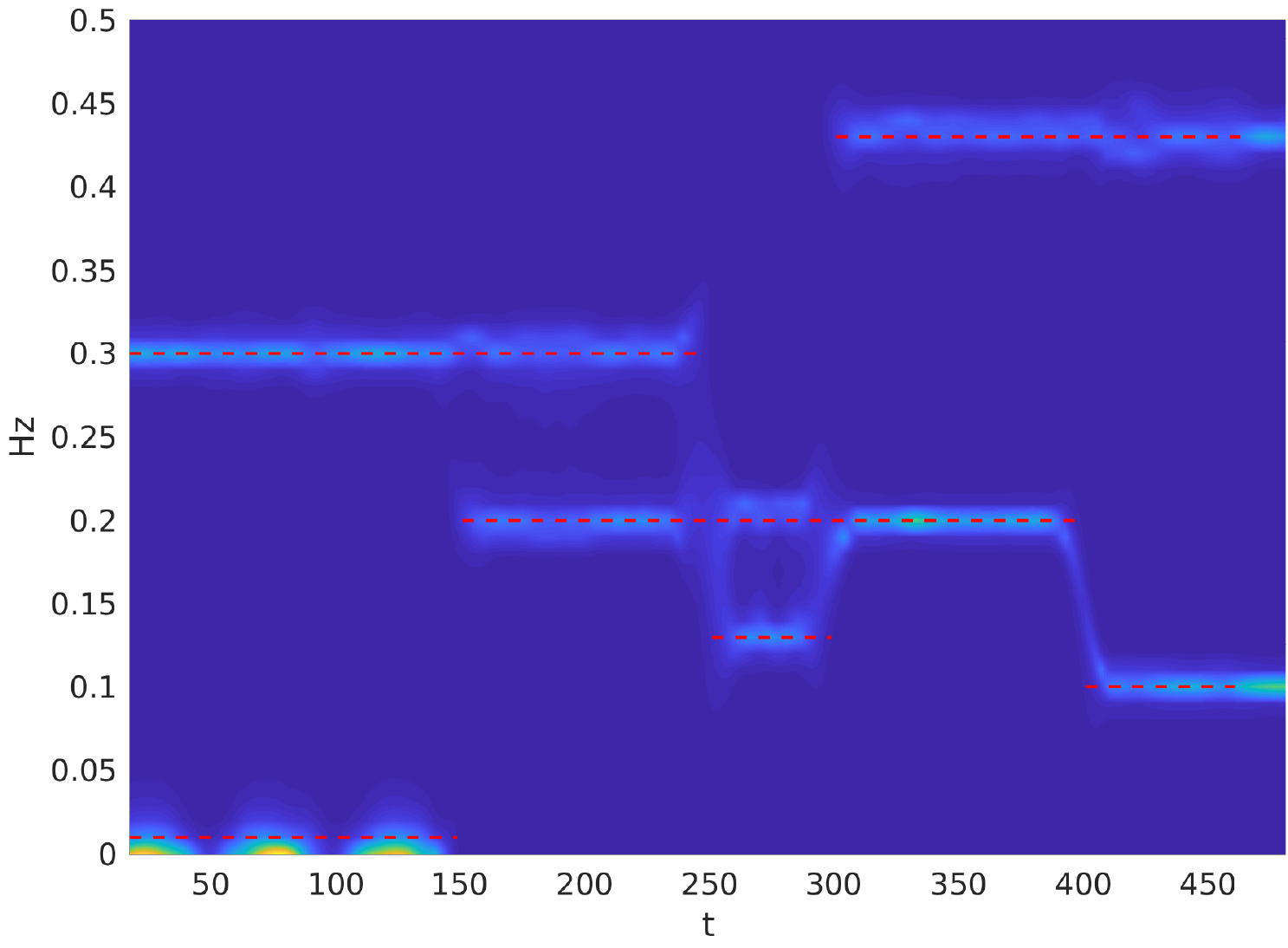}}
	\caption{Spectro-temporal estimation on simulated data. The red dashed lines represent ground truth frequency bands. }
	\label{fig:simu-fig}
\end{figure}

We simulated a noise-observed multi-sinusoidal signal $y(t)$ as shown in \eqref{equ:simu-y} and Fig.~\ref{fig:simu-yt} with time step $\Delta t = 0.1$ and $\varepsilon_k \sim N(0, 0.1^2)$. 

\begin{equation}
\begin{split}
&y(t_k) = \varepsilon_k \\
 &+   \left\{\begin{matrix}
\sin(2\pi \, 0.01 \, t_k) + \sin(2\pi \, 0.3 \, t_k),  &1\leq t_k<150\\ 
\sin(2\pi \, 0.2 \, t_k) + \sin(2\pi \, 0.3 \, t_k),  &150\leq t_k<250\\ 
\sin(2\pi \,0.13 \, t_k) +  \sin(2\pi \, 0.2 \, t_k), &250\leq t_k<300\\
\sin(2\pi \, 0.2 \, t_k) + \sin(2\pi \, 0.43 \, t_k), & 300\leq t_k<400\\ 
\sin(2\pi \, 0.1 \, t_k) + \sin(2\pi \, 0.43 \, t_k), & 400\leq t_k<500
\end{matrix}\right..
\end{split}
\label{equ:simu-y}
\end{equation}
In Fig. \ref{fig:simu-fig}, we plot the time-varying spectrum results using FourierKS, OscKS, STFT, CWT, and BurgAR. The settings for estimation we use here are described in the figure captions. 

Although all methods can approximate the simulated data  
to a good extent, FourierKS and OscKS have higher frequency resolution with less noisy representation which can help us to extract more robust features from spectro-temporal representation. 
Morover, the results from FourierKS and OscKS methods are almost the same although they have different state-space models.

\begin{table}[h!]
	\resizebox{\columnwidth}{!}{
\begin{tabular}{@{}llllll@{}}
		\toprule
		& FourierKS & OscKS & CWT & STFT & BurgAR \\ \midrule
		\begin{tabular}[c]{@{}l@{}}$\Delta t=0.1$ \end{tabular} & 3.39 & 0.18  & 0.08 & 0.07 & 0.36 \\ 
		\begin{tabular}[c]{@{}l@{}}$\Delta t=0.01$ \end{tabular} & 9.18 & 0.95 & 1.32 & 0.30 & 2.58 \\ \bottomrule
	\end{tabular}
}
	\caption{CPU time cost of each spectro-temporal estimation methods. The times are recorded in a MacBook laptop with Core i5 CPU and Matlab 2017b. }
	\label{tab:simu-time}
\end{table}

To verify the computational efficiency of the stationary proposal in Section \ref{sec:osc}, we run each of the estimation methods 20 times and record the mean values of their CPU time. We test with $\Delta t = 0.1$ 
and $\Delta t = 0.01$  
to control the length of the signal. The results in Table~\ref{tab:simu-time} clearly show that the time reduction from FourierKS (3.39~s, 9.18~s) to OscKS (0.18~s, 0.95~s) is significant. For OscKS method, the time for solving DARE is 0.09~s which accounts for almost half of the total time (0.18~s). To reduce the time usage further, one can resort to better DARE solvers or lower resolution in frequency axis. For a longer signal (i.e. $\Delta t = 0.01$), OscKS (0.95~s) method becomes faster than CWT (1.32~s), which indicates a competent efficiency for long signals.

\begin{figure*}[t!]
	\centering
	\includegraphics[width=\linewidth]{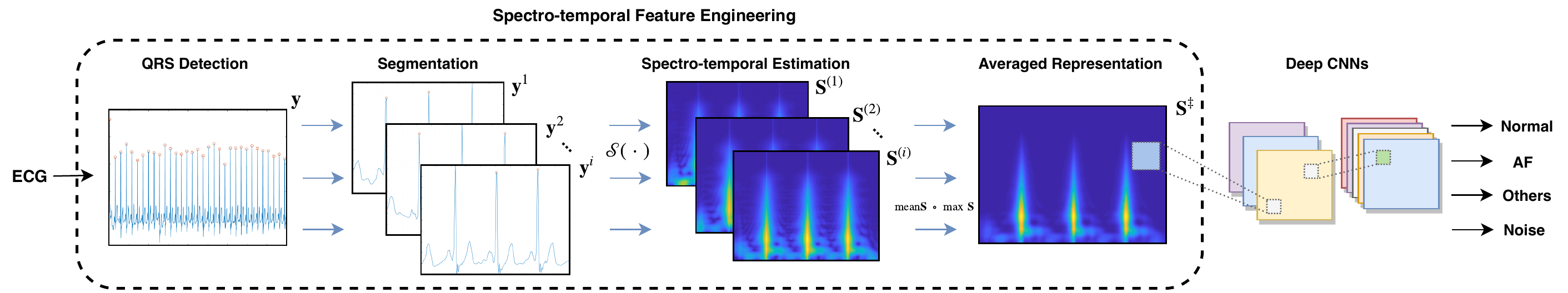}
	\caption{Generalized overall processing scheme for ECG analysis.}
	\label{fig:overall_flowchart}
\end{figure*}

\section{Materials and Methods for ECG Classification}

\subsection{ECG Dataset}\label{new:dataset}

In the AF experiments, we used the ECG dataset provided by PhysioNet/CinC Challenge 2017~\cite{clifford2017af}. In total 8528 short single lead ECG recordings were collected using AliveCor hand-held devices. The recordings were uploaded automatically through an application on the user's mobile phone. In addition, the data were sampled at 300 Hz and band-pass filtered by the AliveCor devices. 
The duration of ECG recordings were between 9~s to 61~s with 30~s median. The distribution of ECG recordings among different classes is as follows: Normal (5076 recordings), AF (758), Other (2415), and Noisy (279). 

\subsection{ECG Spectro-Temporal Feature Engineering}
\label{ssec:ave-overall}

Our aim is now to find the spectro-temporal features of ECG signals such that it can be classified by deep convolutional neural networks (CNNs). In Fig.~\ref{fig:overall_flowchart} we show the overall proposed scheme from input (ECG) to output (predicted label).

The first step is QRS detection and ECG segmentation in which the raw ECG signal is divided into fixed-length segments aligned by their central R peaks. Next, the spectro-temporal data matrix for each segment is calculated using~(\ref{eq:power_density}). The data matrices are then averaged and normalized to generate a fixed-length spectro-temporal feature matrix. In the final step, the 2D feature matrix (spectro-temporal image) is fed into a deep CNN for classification.

\begin{figure}[b!]%
	\centering
	\subfigure[Rec. 5569 (Normal)]{\label{fig:normal}\includegraphics[width=0.45\linewidth]{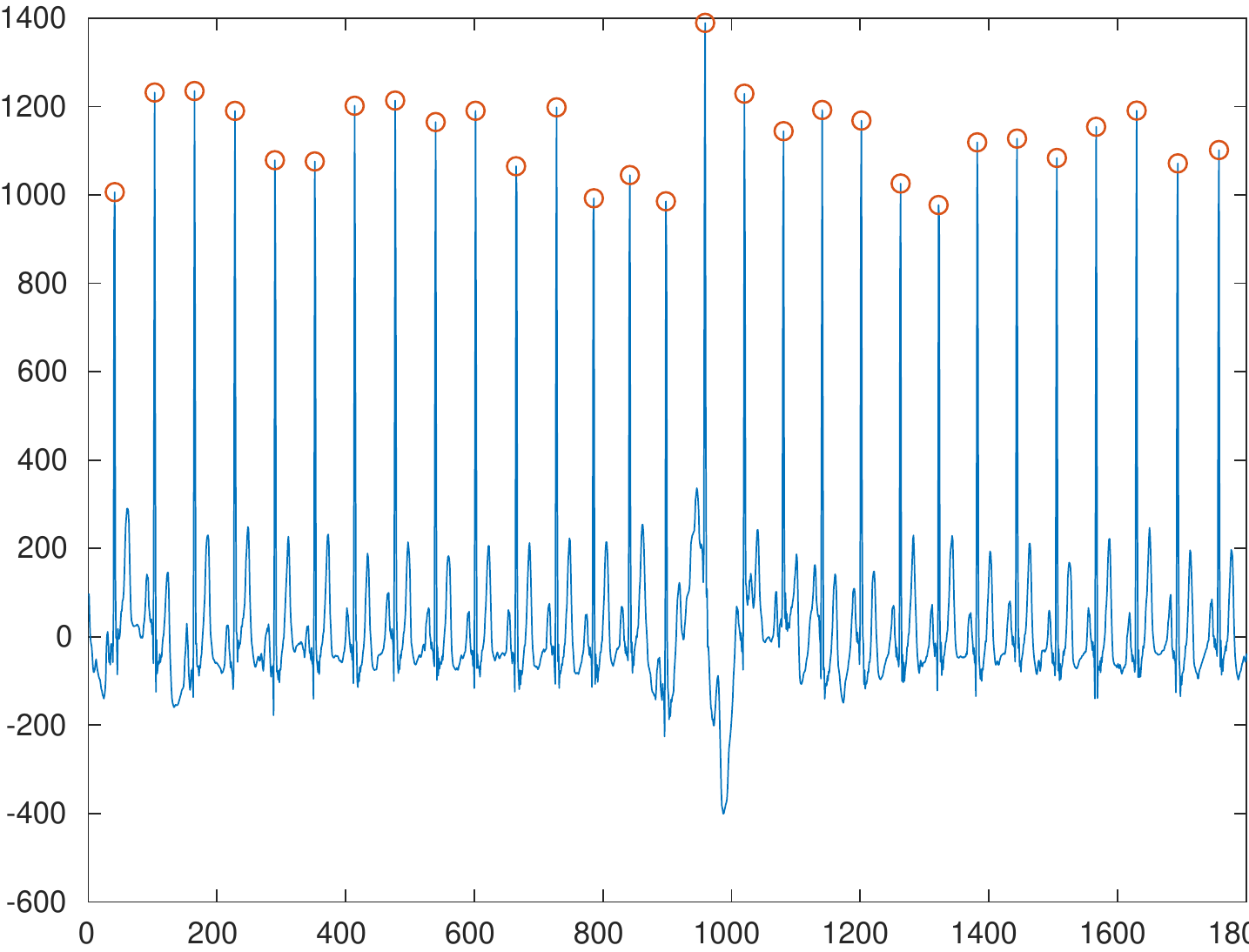}}
	\subfigure[Rec. 5569 (Normal)]{\label{fig:normal}\includegraphics[width=0.45\linewidth]{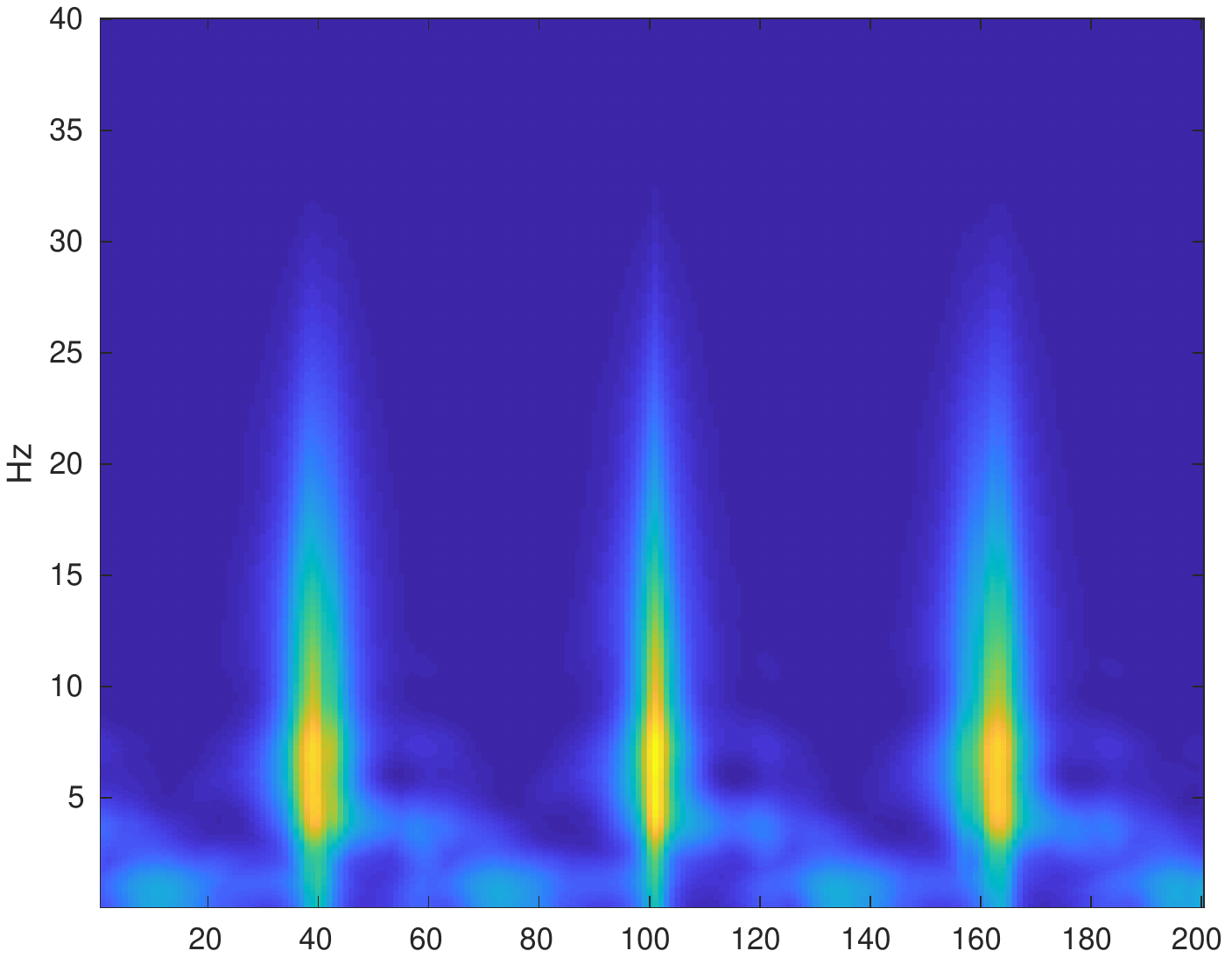}}
	\subfigure[Rec. 5587 (Atrial Fibrillation)]{\label{fig:af}\includegraphics[width=0.44\linewidth]{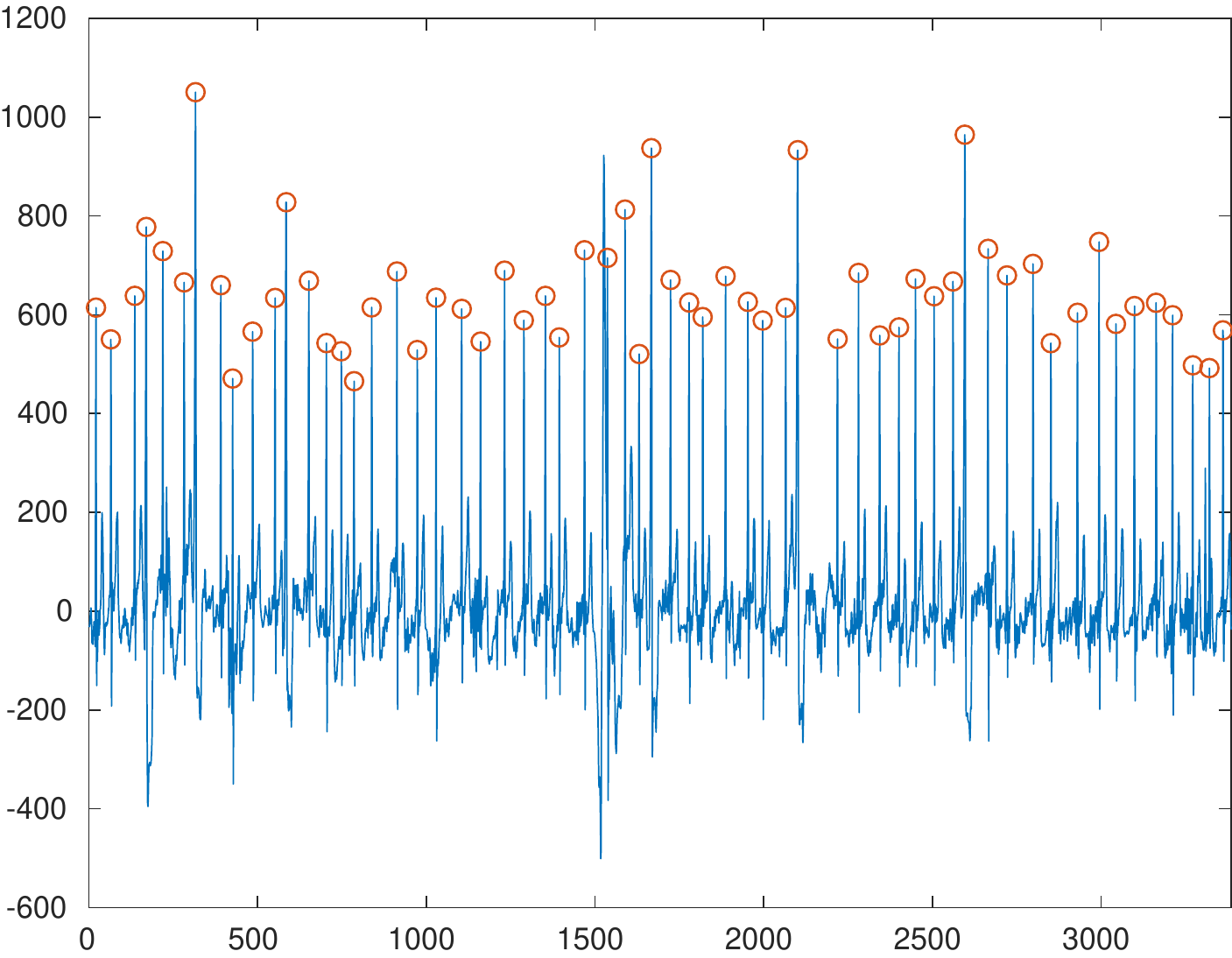}}
	\subfigure[Rec. 5587 (Atrial Fibrillation)]{\label{fig:af}\includegraphics[width=0.45\linewidth]{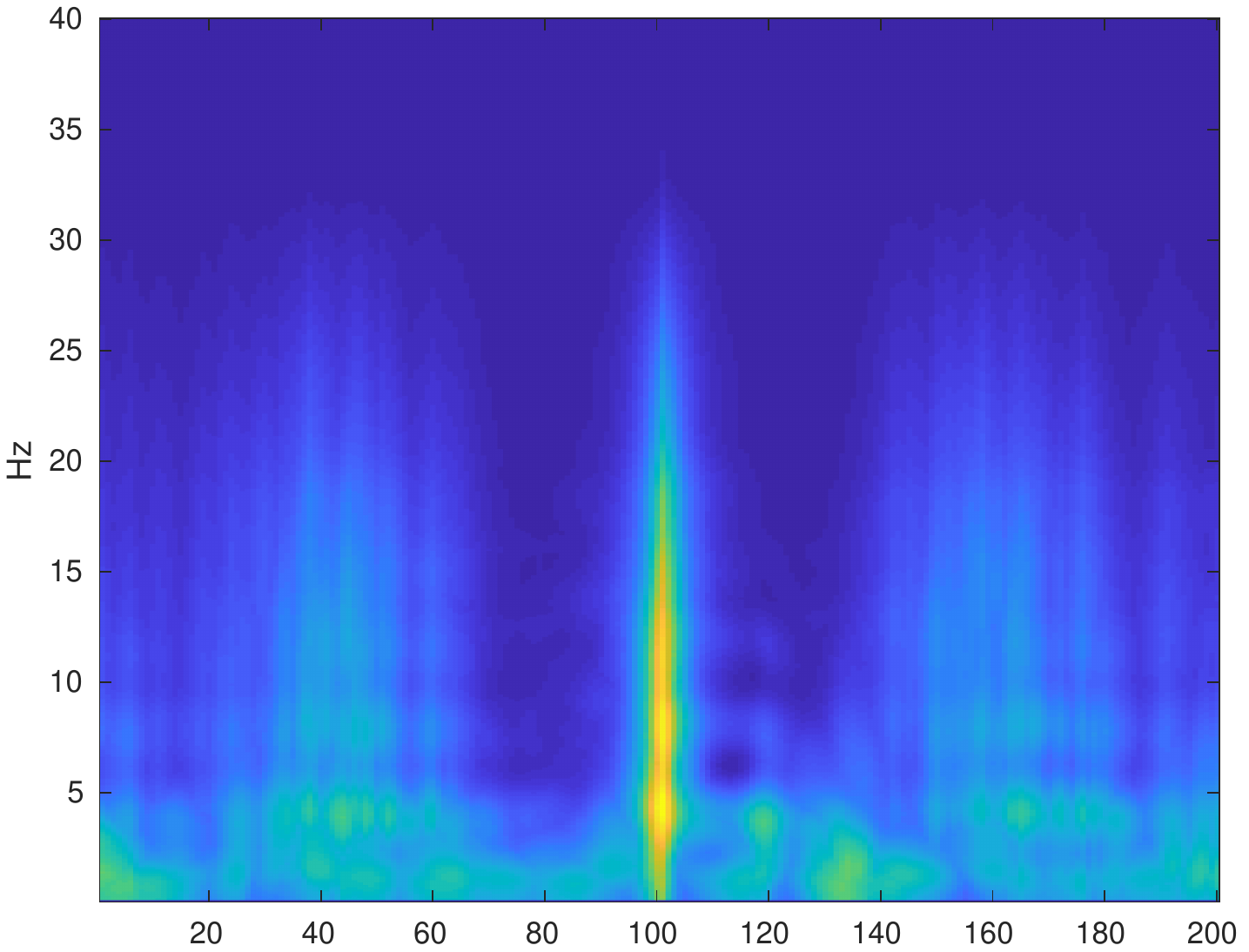}}
	\subfigure[Rec. 5586 (Others)]{\label{fig:other}\includegraphics[width=0.42\linewidth]{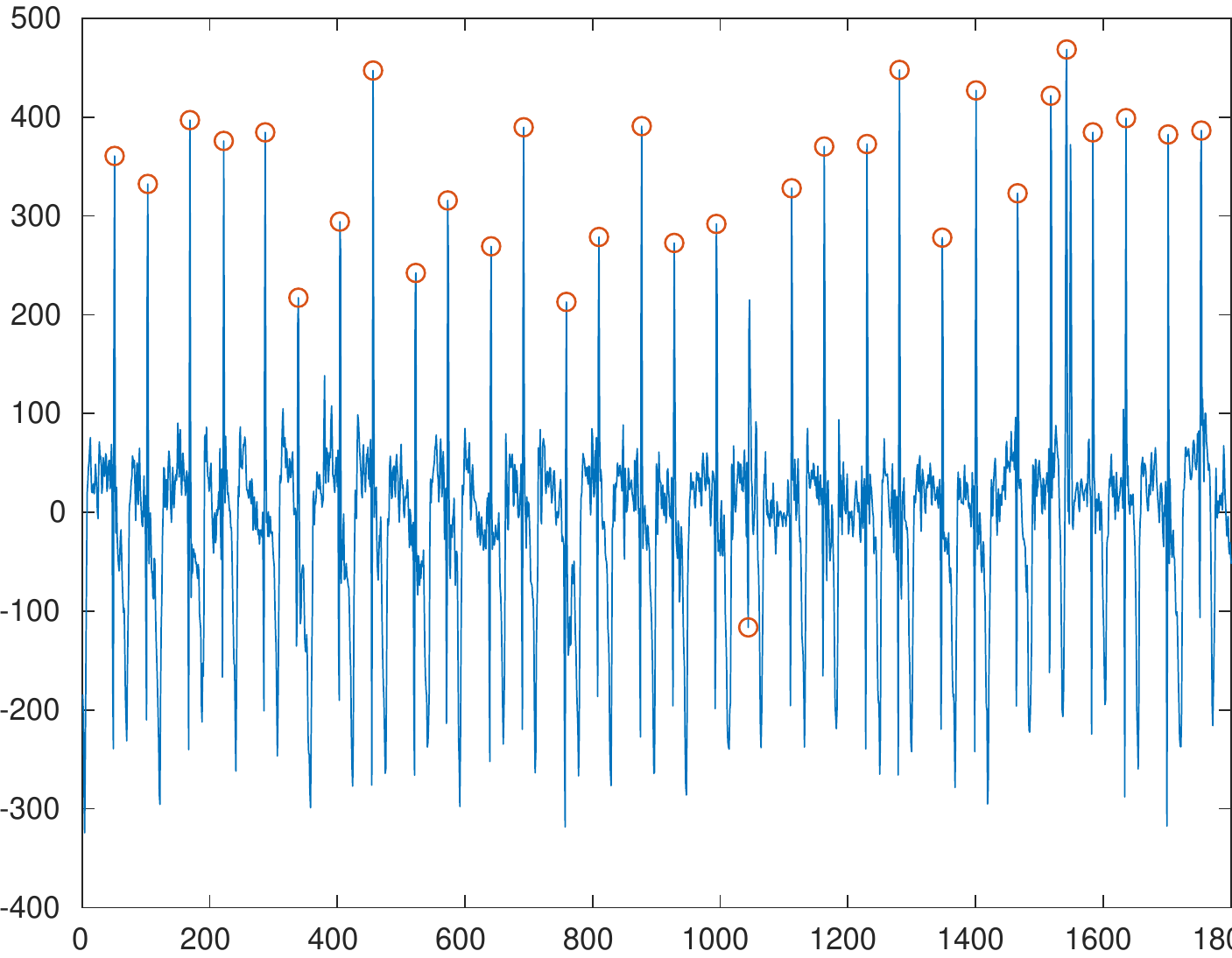}}
	\subfigure[Rec. 5586 (Others)]{\label{fig:other}\includegraphics[width=0.45\linewidth]{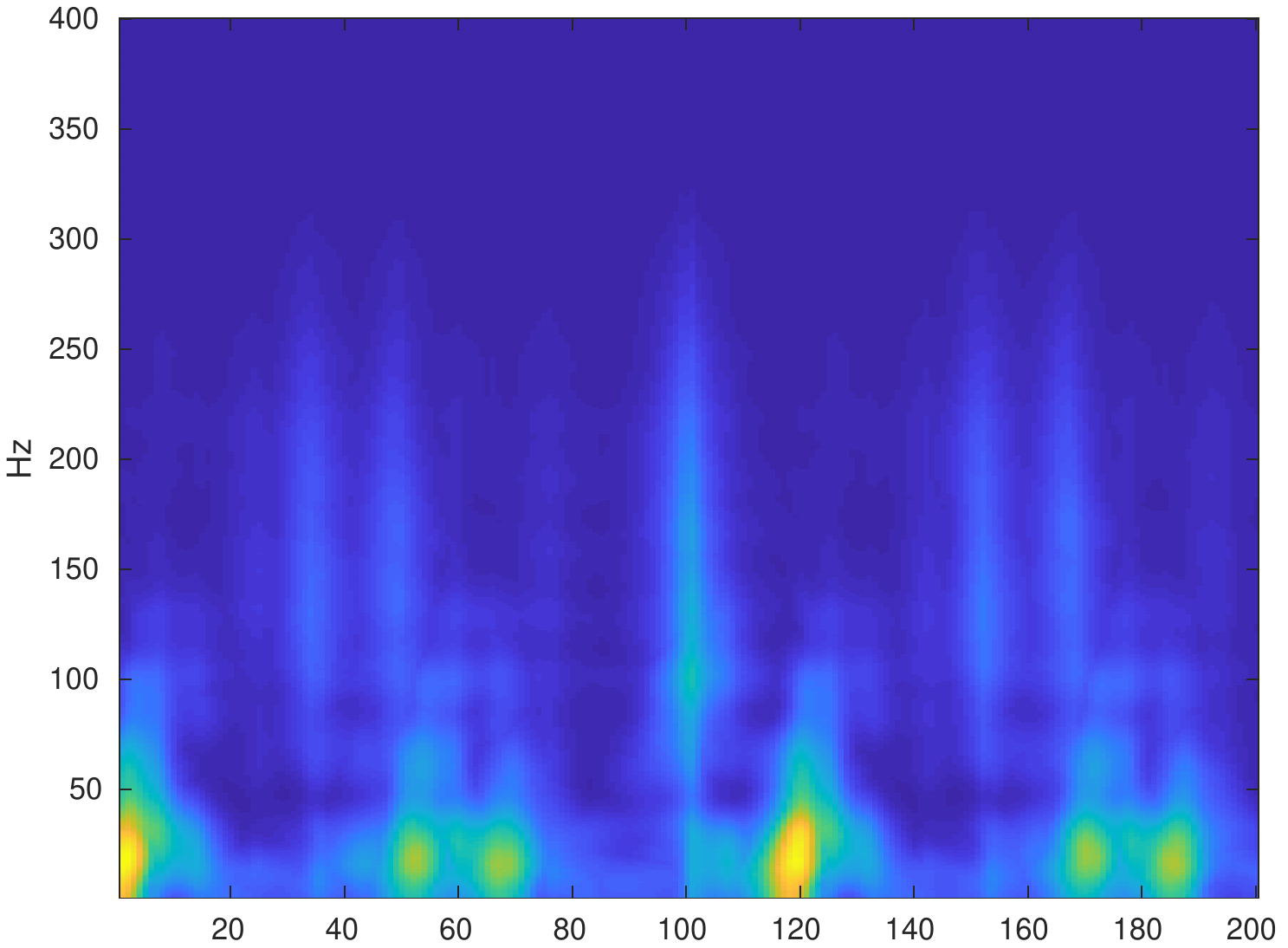}}
	\subfigure[Rec. 5507 (Noise)]{\label{fig:noise}\includegraphics[width=0.43\linewidth]{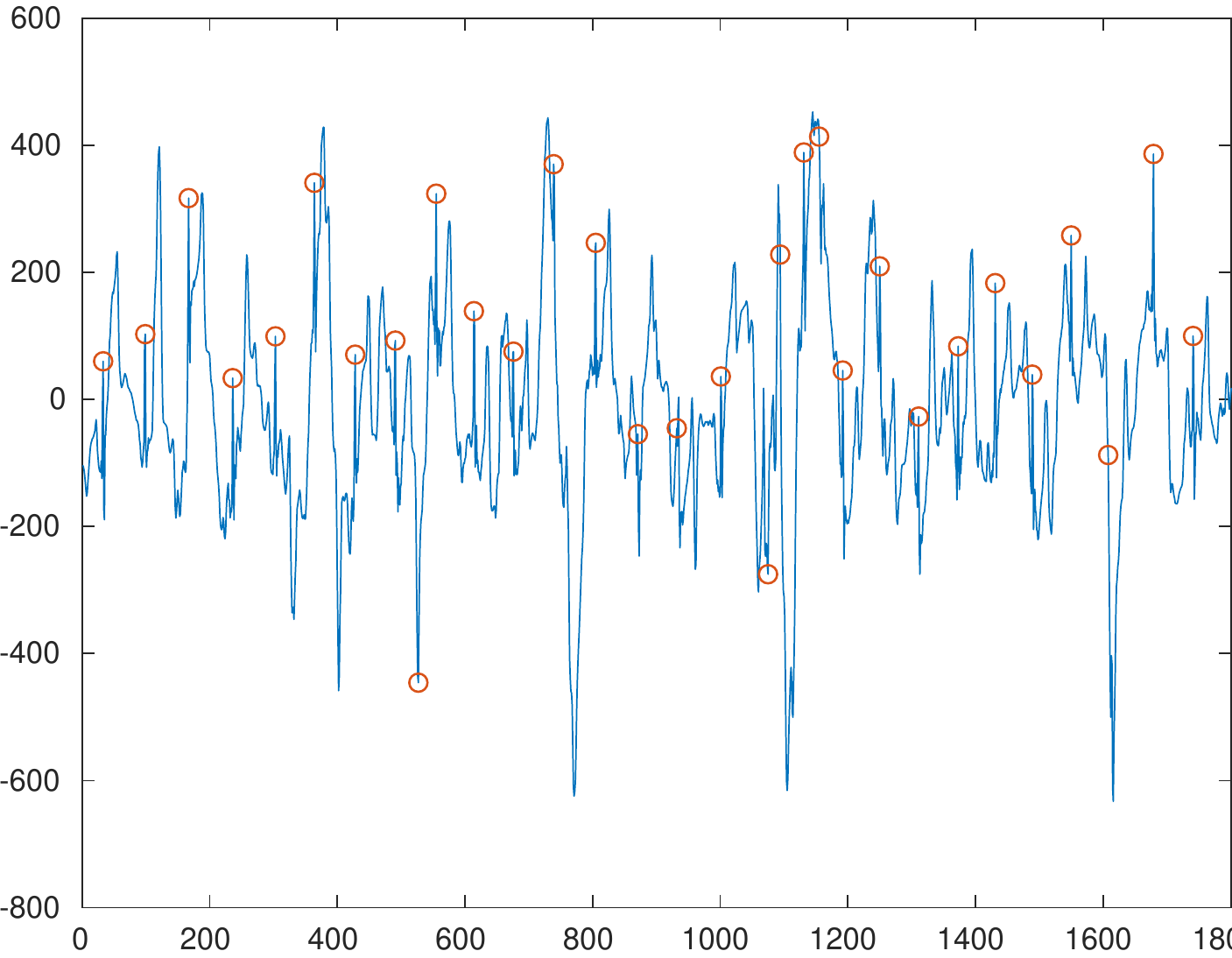}}
	\subfigure[Rec. 5507 (Noise)]{\label{fig:noise}\includegraphics[width=0.455\linewidth]{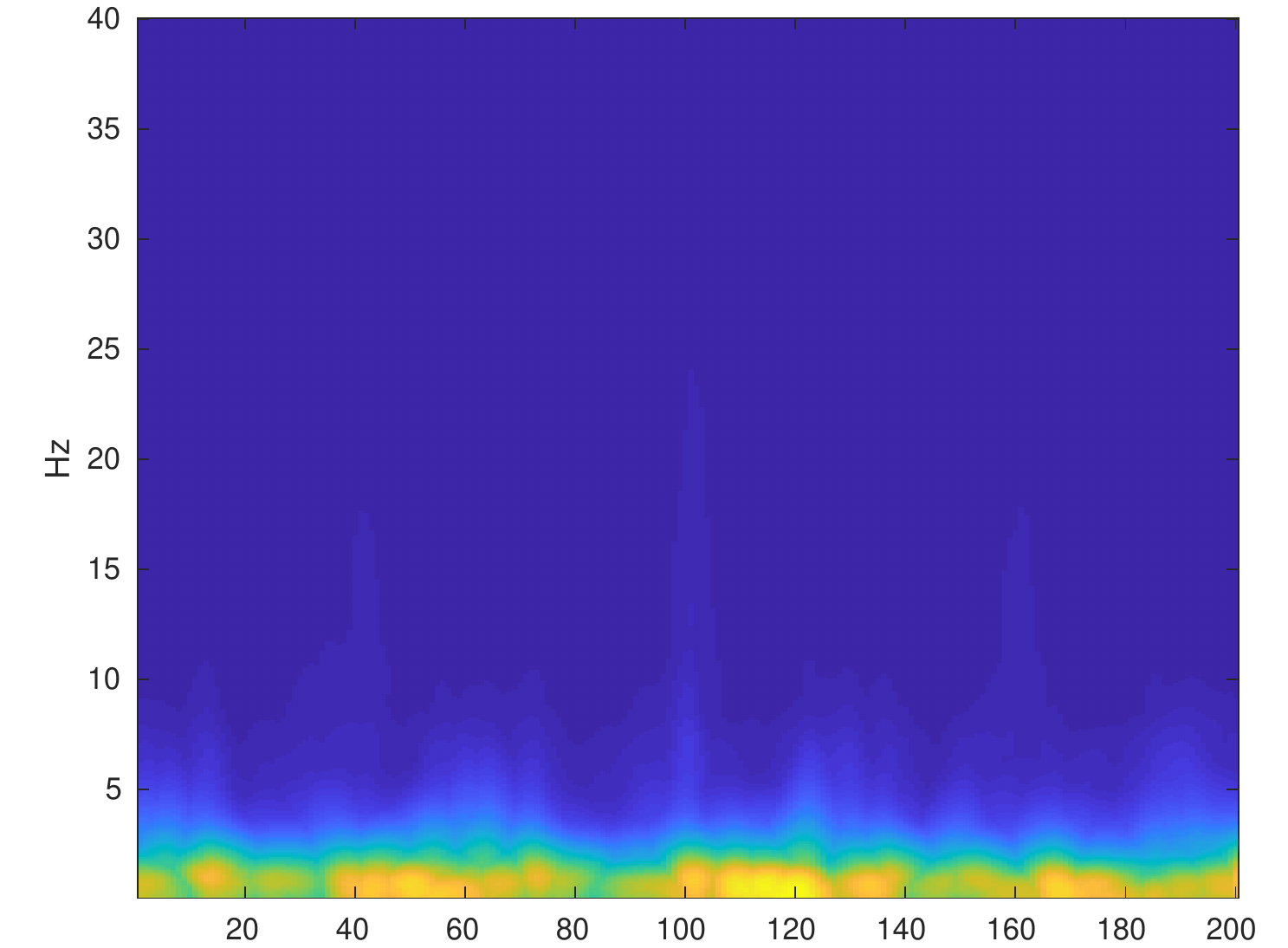}}
	\caption{Results of representation averaging (right side) on four types of ECG signals (left side), using proposed spectro-temporal method. Red circles indicate detected R peaks.}
	\label{fig:pre_ave_st_show}
\end{figure}

The logic behind the segmentation and averaging steps in the feature engineering procedure (dashed area in Fig.~\ref{fig:overall_flowchart}) is threefold. First, it can handle the problem of ECG recordings with different length, and generate fixed-length spectro-temporal feature matrices. Second, it can capture enough information from ECG recording to be classified by CNNs. For example, since the central R peaks in each segments are aligned, after averaging we expect sharp edges corresponding to QRS complexes in feature matrices (spectro-temporal image) for Normal rhythms. However, for AF rhythms we expect the blurred area in spectro-temporal images due to the variable R-R intervals. For, noisy segments we do not expect any clear area for QRS complexes, and for Other classes based on the underlying arrhythmia one can expect different patterns in spectro-temporal images (see Fig.\ref{fig:pre_ave_st_show}). Finally, the third reason to use the segmentation and averaging steps is to decrease the effect of noise in ECG recordings. In the following we discuss different steps of feature engineering in detail.

\begin{figure}[htp!]
	\centering
	\subfigure[QRS detection using Pan-Tompkin]{\label{fig:pt-org}\includegraphics[width=0.45\linewidth]{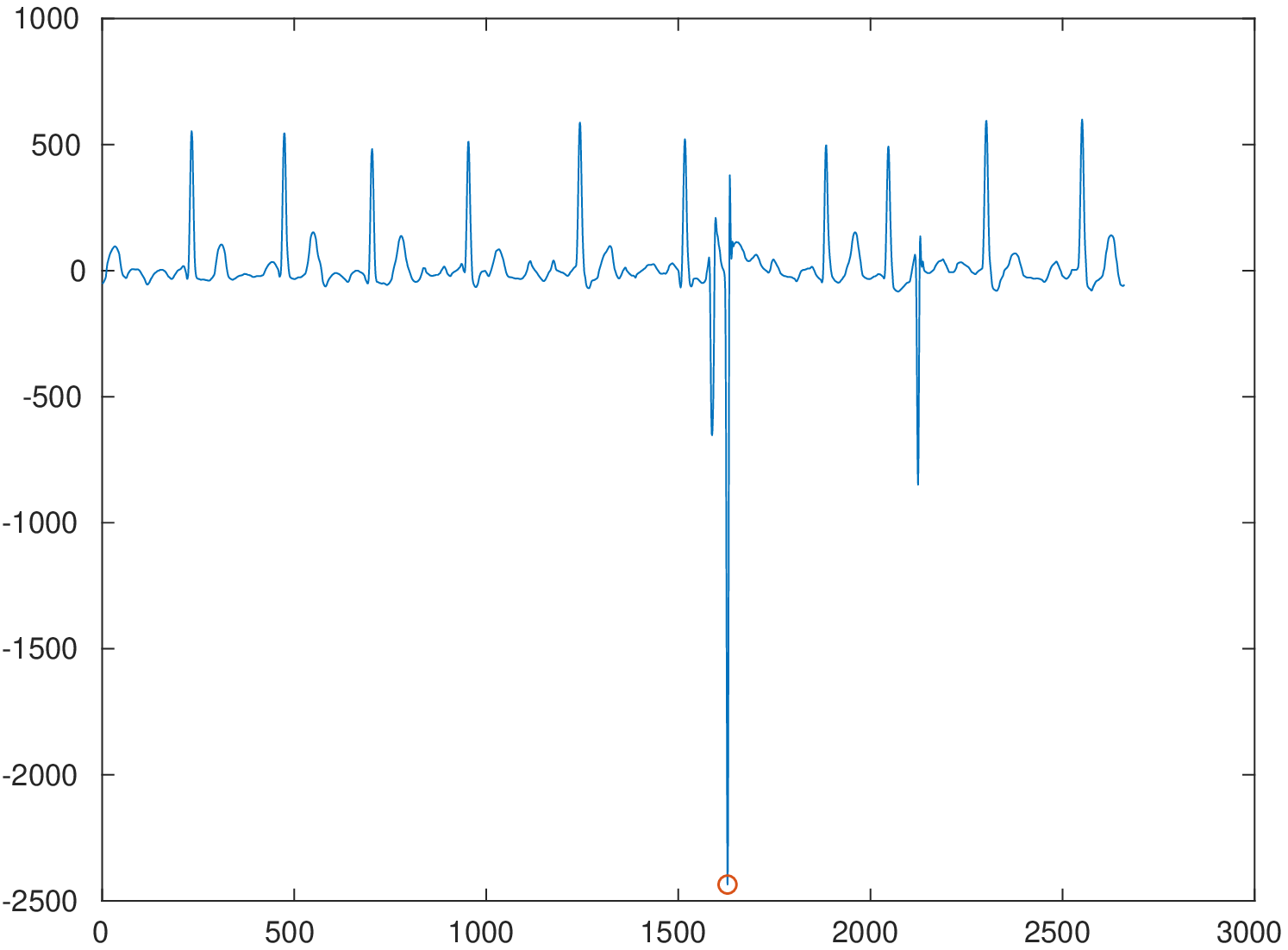}}
	\subfigure[QRS detection using iterative Pan-Tompkin]{\label{fig:pt-rev}\includegraphics[width=0.45\linewidth]{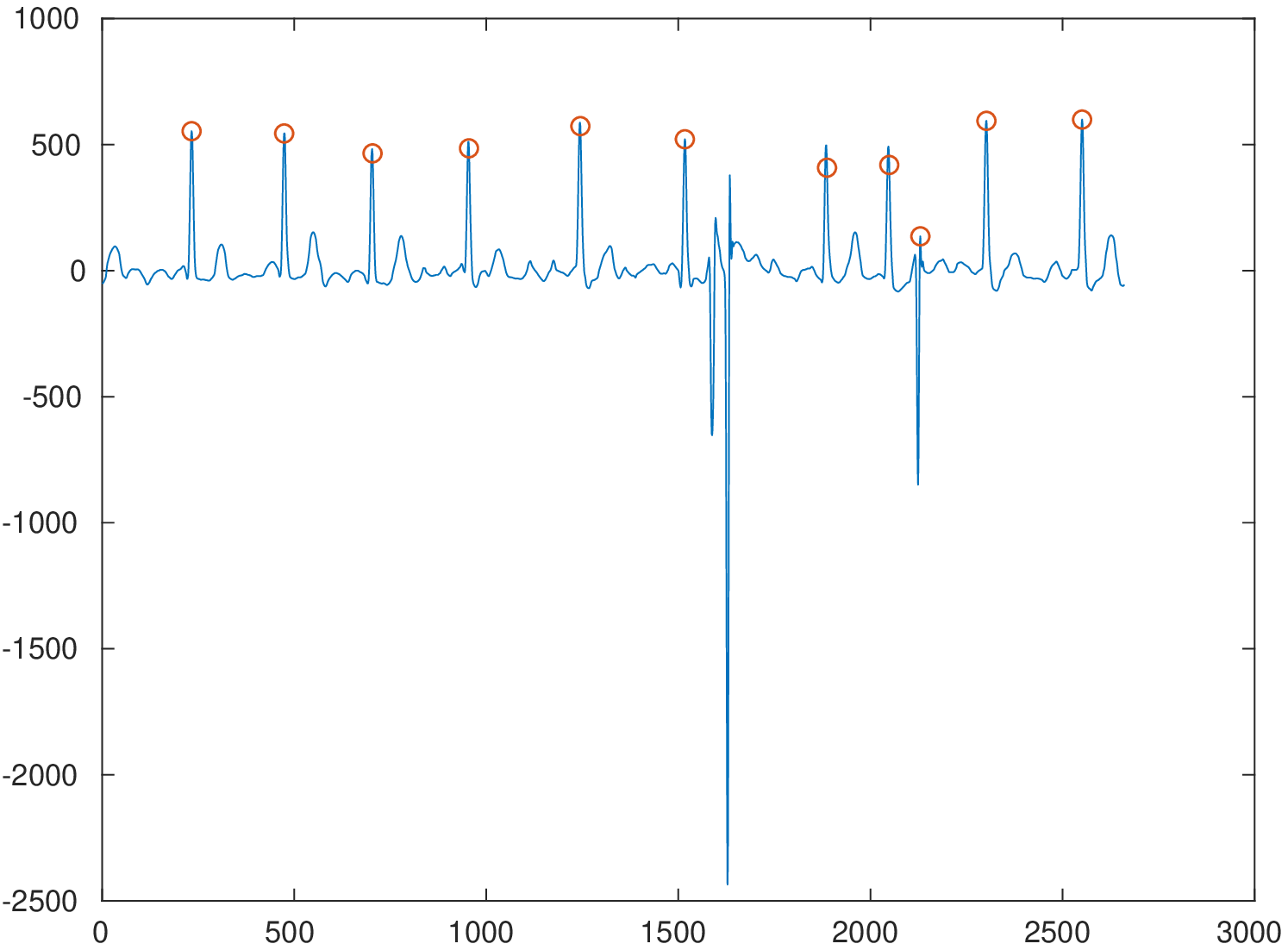}}
	\caption{Improvement in QRS detection}
	\label{fig:pt-diff}
\end{figure}

In this work, for QRS detection, we use a modified version of Pan-Tompkins algorithm. The original Pan-Tompkins algorithm~\cite{pan1985real} is sensitive to burst noise, and it easily misinterprets noise with R peak. To address this limitation at least partially, we slightly modify the original algorithm such that it iteratively checks the number of detected R peaks and if that number is smaller than a threshold, it ignores the detected R peaks and their neighbourhood samples in the ECG signal, and again applies the Pan-Tompkins algorithm on the rest of the signal. In this way, if there are few instances with high-amplitude burst noise, our algorithms can handle those. One example which illustrate this modification is shown in Fig.~\ref{fig:pt-diff}.

The next step is segmentation in which the fixed-length ECG segments are extracted from the original signal such that each segment potentially covers three QRS complexes. The segmentation process is described as follows: if $\mathbf{y} = \begin{bmatrix} y_1& y_2& \cdots & y_N \end{bmatrix}^\top \in \mathbb{R}^{N}$ is the original ECG signal and $\bar{p}_i \in \{1,2,\cdots,N\}$ is the position of $i$th R peak in $\mathbf{y}$, then $\mathbf{\bar{p}} =\begin{bmatrix} \bar{p}_1 & \bar{p}_2 & \cdots & \bar{p}_D \end{bmatrix}^\top$ holds the positions of all R peaks, and $D$ is the total number of R peaks in $\mathbf{y}$. Now, to extract $D-2$ ECG segments we associate each $\bar{p}_i$, $i \in \{2,\cdots,D-1 \}$, to a segment of $\mathbf{y}$ such that it potentially covers three adjacent QRS complexes. To do so, we collect $\beta$ samples before and after each $\bar{p}_i$. Following this procedure, the ECG segment associated to $i$th R peak can be extracted from $\mathbf{y}$ as $\mathbf{y}^{(i)} = \begin{bmatrix} {y}_{\bar{p}_i - \beta} & \cdots & {y}_{\bar{p}_i} & \cdots& {y}_{\bar{p}_i + \beta} \end{bmatrix}^\top$, and using equation (\ref{eq:power_density}), the spectro-temporal data matrix corresponding to this ECG segment is $\mathbf{S}^{(i)}\in\mathbb{R}^{M\times (2\beta+1)}$ where $M$ and $2\beta+1$ are frequency and time steps, respectively. It is worth noticing that these two parameters (i.e., $M$ and $2\beta+1$) determine the size of the matrix $\mathbf{S}$ in~(\ref{eq:power_density}). The choice of parameter $\beta$ is important, as it regulates the length of output and how much takes into average. 
Usually, $\beta$ should cover at least three QRS complexes for good evidence of R-R intervals. 

\begin{algorithm}[h!]
	\caption{Averaging representation}\label{psc:pre-pro}
	\begin{algorithmic}[1]
		\Input Signal $\cu{y} = \left [ y_1, y_2, ..., y_N \right ]^\top$
		\Output Spectro-temporal feature $\mathbf{S}^{\ddagger}\in\mathbb{R}^{M\times (2\beta+1)}$
		\State Perform Pan-Tompkins on $\cu{y}$ and obtain $\mathbf{\bar{p}}$
		\State $D = \text{size}(\mathbf{\bar{p}})$
		\If {$D \leq \delta$}
			\State $\bar{\cu{y}} = \cu{y}$
			\State $\bar{\cu{y}}_{i-\alpha:i+\alpha}=0$ for all $i \in \mathbf{\bar{p}}$
			\State Perform Pan-Tompkins on $\bar{\cu{y}}$, and obtain new $\mathbf{\bar{p}}, D$
		\EndIf
		\ForAll {$1< i<D $ in $\mathbf{\bar{p}}$}
			\State Perform Spectro-temporal estimation on $\cu{y}^{(i)}$ to get $\cu{S}^{(i)}$
		\EndFor\\
		\Return $\mathbf{S}^{\ddagger} = \frac{\sum^{D-1}_{i=2} \mathbf{S}^{(i)}}{D-2} \circ \max_{2\leq i \leq D-1} \mathbf{S}^{(i)}$
	\end{algorithmic}
\end{algorithm}

The spectro-temporal feature matrix $\mathbf{S}^{\ddagger}$ is obtained by averaging over all spectro-temporal data matrices and multiplying with their maximum mask: 
\begin{equation}
\mathbf{S}^{\ddagger} = \frac{\sum^{D-1}_{i=2} \mathbf{S}^{(i)}}{D-2} \circ \max_{2\leq i \leq D-1} \mathbf{S}^{(i)}.
\label{equ:average}
\end{equation}

The reason for adding a $\max$ operation in Equation \eqref{equ:average} is that it could, at least in certain extent, help preserving intricate details of spectro-temporal data that were potentially lost during averaging across every segments, and also normalizing the data.

Examples of ECG spectro-temporal feature matrices (images) four different classes of ECG signals are shown in Fig.~\ref{fig:pre_ave_st_show}, where we used the proposed spectro-temporal estimation method in Section \ref{sec:osc}.

\subsection{Classification}

\begin{figure}[htb!]
	\centering
	\includegraphics[width=0.8\linewidth]{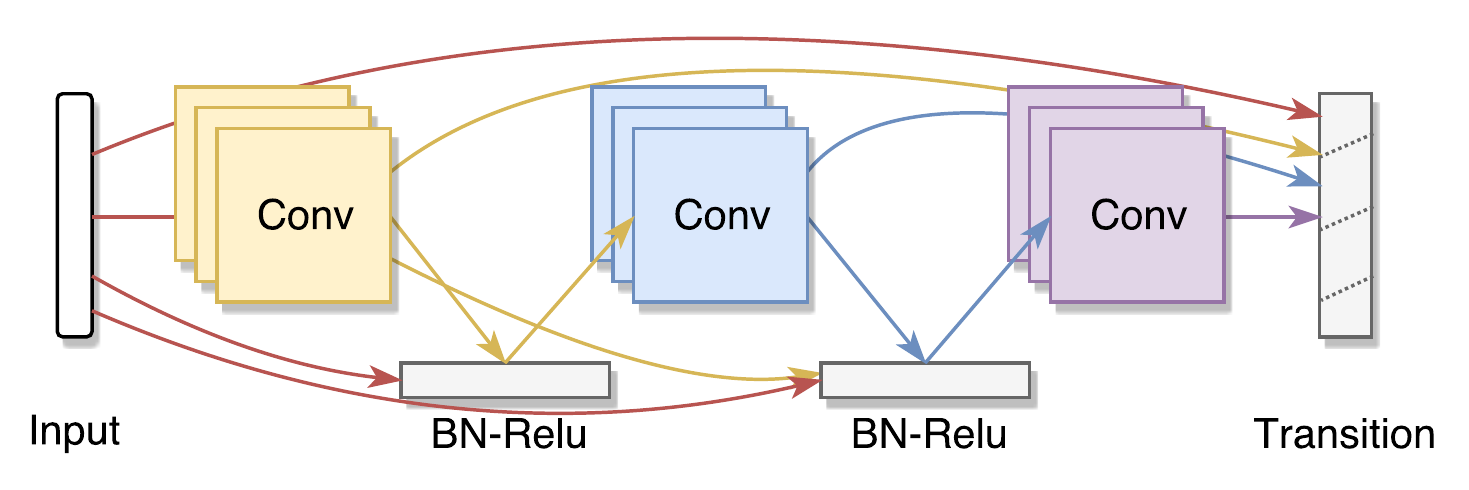}
	\caption{Dense block: each of of the convolutional layer takes all of their preceding outputs as input.}
	\label{fig:dense_block}
\end{figure}
In the recent ten years, deep learning techniques, especially convolutional neural networks, have achieved great success in detection and classification tasks. Comparing to 1D CNNs models, the progress of CNNs for 2D image applications is more prosperous. The aim here is to leverage advanced CNNs for AF classification using the time-varying spectrum (which is an image).

However, one flaw in most of the current network models is that the information during training, principally the gradient, may disappear if the network is exceedingly deep (with many layers), which is usually called ``vanishing gradient'' \cite{glorot2010understanding}. In general way, this root problem can be alleviated by several basic ways, for instance, with pre-training, residual connection, or with properly selected activation functions (e.g., one should not attach ReLu before batch normalization).

Densely connected convolutional networks (DenseNet) \cite{huang2017densely}, which won the 2017 best paper award of CVPR, provide state-of-the-art performance without degradation or over-fitting even when stacked by hundred of layers. DenseNets can be seen as refined versions of deep residual networks (ResNets) \cite{he2016deep}, where the former one introduces explicit connection on every two and preceding layers in a dense block rather than only adjacent layers, as shown in Fig.~\ref{fig:dense_block}. Another additional advantage of DenseNet, as mentioned in \cite{huang2017densely}, is the feature reuse. 

Considering an $L$ layers network, and image input $\mathbf{U}_0$, the output of $l$-th layer is:
\begin{gather}
\mathbf{U}_l = H_l^{Res}(\mathbf{U}_{l-1}) + \mathbf{U}_{l-1}, \\
\mathbf{U}_l = H_l^{Den}(\begin{bmatrix} \mathbf{U}_{0} & \mathbf{U}_{1} &\cdots & \mathbf{U}_{l-1}\end{bmatrix}). 
\label{equ:densenet}
\end{gather}
where $H_l^{Res}$ and $H_l^{Den}$ are layer operations (e.g., convolution, batch-normalization, or activation) of ResNet and DenseNet respectively, and $\mathbf{U}_l$ is the output of $l$th layer.

The DenseNet we implement here, which we refer as Dense18$^+$, is slightly different from the original proposal \cite{huang2017densely}, where we employ both max and average global pooling on last layer and concatenate them as shown in Table~\ref{tbl:dense-stru}. In our application, because of the size of input, we remove the initial down-sampling max pooling layer. Each dense block contains four $3 \times 3$ convolutional layers, with growth rate of 48 and reduction rate 0.5. 
\begin{table}[htb!]
	\centering
	\begin{tabular}{@{}lcc@{}}
		\toprule
		\textbf{Layer Name} & \textbf{Structure} & \textbf{Output Size} \\ \midrule
		Input & Input & (50, 50, 1) \\ \midrule
		Convolution & \begin{tabular}[c]{@{}c@{}}$7\times 7$ conv\\ stride 1\end{tabular} & (50, 50, 64) \\ \midrule
		Dense Block 1 & $\begin{bmatrix}
		1\times 1 & \text{conv}\\ 
		3\times 3 & \text{conv}
		\end{bmatrix} \times 4$ & (50, 50, 256) \\ \midrule
		Transition 1 & \begin{tabular}[c]{@{}c@{}}$1\times 1$ conv\\ $2\times 2$ ave pool\end{tabular} & (25, 25, 128) \\ \midrule
		Dense Block 2 & $\begin{bmatrix}
		1\times 1 & \text{conv}\\ 
		3\times 3 & \text{conv}
		\end{bmatrix} \times 4$ & (25, 25, 320) \\ \midrule
		Transition 2 & \begin{tabular}[c]{@{}c@{}}$1\times 1$ conv\\ $2\times 2$ ave pool\end{tabular} & (12, 12, 160) \\ \midrule
		Dense Block 3 & $\begin{bmatrix}
		1\times 1 & \text{conv}\\ 
		3\times 3 & \text{conv}
		\end{bmatrix} \times 4$ & (12, 12, 352) \\ \midrule
		Transition 3 & \begin{tabular}[c]{@{}c@{}}$1\times 1$ conv\\ $2\times 2$ ave pool\end{tabular} & (6, 6, 176) \\ \midrule
		Dense Block 4 & $\begin{bmatrix}
		1\times 1 & \text{conv}\\ 
		3\times 3 & \text{conv}
		\end{bmatrix} \times 4$ & (6, 6, 368) \\ \midrule
		\begin{tabular}[c]{@{}l@{}}Pooling\\ Concatenate\end{tabular} & $\begin{bmatrix}
			\text{global ave}\\
			\text{global max}
		\end{bmatrix} \text{concat}$ & (736) \\ \midrule
		\begin{tabular}[c]{@{}l@{}}Fully Connected\\ (Softmax)\end{tabular} & 4 classes & (4) \\ \bottomrule
	\end{tabular}
	\caption{Structure of Dense18$^+$ in this paper.}
	\label{tbl:dense-stru}
\end{table}

\begin{table*}[t!]
	\centering
	\resizebox{\linewidth}{!}{
	\begin{tabular}{@{}lccccccc@{}}
		\toprule
		$F1_{overall}$ & \multicolumn{1}{l}{Random Forest\cite{liaw2002classification}} & \multicolumn{1}{l}{CNN18} & \multicolumn{1}{l}{InceptionV3\cite{Szegedy_2016_CVPR}} & \multicolumn{1}{l}{ResNet18\cite{he2016deep}} & \multicolumn{1}{l}{ResNet34\cite{he2016deep}} & \multicolumn{1}{l}{DenseNet18\cite{huang2017densely}} & \multicolumn{1}{l}{Dense18$^+$} \\ \midrule
		STFT & 73.47 & 72.65 & 75.66 & 76.17 & 76.26 & 77.39 & \textit{77.67} \\
		CWT & 74.91 & \textbf{73.96} & 76.41 & \textbf{78.57} & \textbf{78.70} & 78.82 & \textit{79.63} \\
		BurgAR & 73.22 & 71.78 & 76.45 & 76.41 & 76.30 & 77.58 & \textit{77.76} \\
		FourierKS & 75.99 & 72.74 & \textbf{77.48} & 78.05 & 77.99 & 79.50 & \textit{\textbf{80.24}} \\
		OscKS & \textbf{76.12} & 73.07 & 76.91 & 77.85 & 78.19 & \textbf{79.67} & \textit{80.18} \\ \bottomrule
	\end{tabular}
}
	\caption{10-fold cross-validation F1 Score of spectro-temporal estimation methods using different classifiers for classification. Best score for each column and row are rendered bold and italic respectively. }
	\label{tbl:f1-overall}
\end{table*}

\begin{table*}[t!]
	\centering
	\begin{tabular}{@{}llllllll@{}}
		\toprule
		& Method & $F1_N$ & $F1_A$ & $F1_O$ & $F1_\sim$ & $F1_{overall}$ & $Std_{F1}$ \\ \midrule
		(1) & STFT + Dense18$^+$ & 88.67 & 74.49 & 69.84 & 53.28 & 77.67 & 1.78 \\
		(2) & CWT + Dense18$^+$ & \textbf{89.30} & 77.76 & 71.82 & 51.95 & 79.63 & 1.76 \\
		(3) & BurgAR + Dense18$^+$ & 88.35 & 75.17 & 69.74 & \textbf{56.49} & 77.76 & 1.62 \\
		(4) & Kalman + Dense18$^+$ & 89.29 & 79.18 & \textbf{72.25} & 52.50 & \textbf{80.24} & \textbf{1.52} \\
		(5) & OSC + Dense18$^+$ & 89.09 & \textbf{79.78} & 71.68 & 55.86 & 80.18 & 1.55 \\ 
		(6) & Martin \cite{zihlmann2017convolutional} & 88.8 & 76.4 & 72.6 & 64.5 & 79.2 & N/A \\
		(6) & Zhaohan \cite{xiong2017robust} & 87 & 80 & 68 & N/A & 78 & N/A \\ \bottomrule& 
	\end{tabular}
	\caption{10-fold cross-validation results of overall and four labels using different spectro-temporal estimation methods on Dense18$^+$ classifier. Best score for each column are rendered bold.}
	\label{tbl:detail-compare}
\end{table*}

\subsection{Model Assessment and Evaluation Criteria}\label{ssec:dataset}

To evaluate the performance of the proposed methods, we have conducted experiments on the ECG dataset described in Section~\ref{new:dataset}. 
The classification performance of different methods was assessed by using the scoring mechanism recommended by PhysioNet/Computing in Cardiology (CinC) Challenge 2017~\cite{clifford2017af} over the whole dataset in 10-fold cross-validation scheme. The data were partitioned such that the same proportions of each class are available in each fold (stratified cross-validation). Moreover, the F1 score,
\begin{equation}
F1 = 2 \cdot \frac{\textnormal{Precision} \cdot \textnormal{Recall}}{\textnormal{Precision}+\textnormal{Recall}}
\end{equation}
for each class is calculated to summarize the performance of that specific class: Normal ($F1_N$), AF ($F1_A$), Others ($F1_O$), and Noisy ($F1_\sim$). Then, as recommended by PhysioNet/CinC 2017 the overall evaluation metric is used as follows:
\begin{equation}
F1_{\textnormal{overall}} = \frac{1}{3}(F1_N + F1_A + F1_O). 
\end{equation}
Finally, the detailed performance is shown by a 4-class confusion matrix whose the diagonal entries are the correct classifications and the off-diagonal entries are the incorrect classifications. This confusion matrix is the result of stacking 10 confusion matrices of the test data in the 10-fold cross-validation.
 
\section{Experiments}\label{results1}

In principle, any time-frequency analysis 
method can be used 
for ECG classification. So, in order to show the benefit of using the proposed spectro-temporal method 
in Section \ref{sec:st-overall} over other 
standard time-frequency analysis methods, we have conducted experiments on the ECG dataset. We have compared the results of the proposed method with short-time Fourier transform (STFT), continuous wavelet transform (CWT), and classical power spectral density estimation method. To do so, we used magnitude of STFT, magnitude of CWT, and square root of non-logarithmic power spectral density using Burg autoregressive model (BurgAR)~\cite{kay1981spectrum} of ECG signal to construct the feature matrices.

In addition, several different convolutional architectures are examined, and their results are compared to the standard RF classifier.
The networks structure of InceptionV3, ResNet, and DenseNet are taken from their original papers~\cite{Szegedy_2016_CVPR, he2016deep, huang2017densely}, but we removed the initial sub-sampling layer for a fair comparison with Dense18$^+$ in Table~\ref{tbl:dense-stru}. We also construct a plain CNN (CNN18) which has the same structure setting with Dense18$^+$ but without dense connection. For the random forest we use 500 decision trees and random selection of 50 features (out of 2500) at each node. In addition, at each node the random forest minimizes the cross-entropy impurity measure. The settings for spectro-temporal estimation we choose here are the same as described in Section \ref{ssec:time-freq}. All spectro-temporal feature matrices (images) are then unifiedly resized (down-sample by local averaging) to $50\times 50$ for classifiers.

With seven classifiers and five different time-frequency analysis methods, in total we have 35 different combinations whose performance are reported in Table~\ref{tbl:f1-overall}. As can be seen from this table the best results (overall scores) belong to our proposed spectro-temporal representation methods (i.e., FourierKS and OscKS) with Dense18$^+$ classifier. Moreover, Table~\ref{tbl:detail-compare} shows the performance for each ECG classes for Dense18$^+$ classifier with different time-frequency representation.

\begin{figure*}[tb!]%
	\centering
	\subfigure{\includegraphics[width=0.3\linewidth]{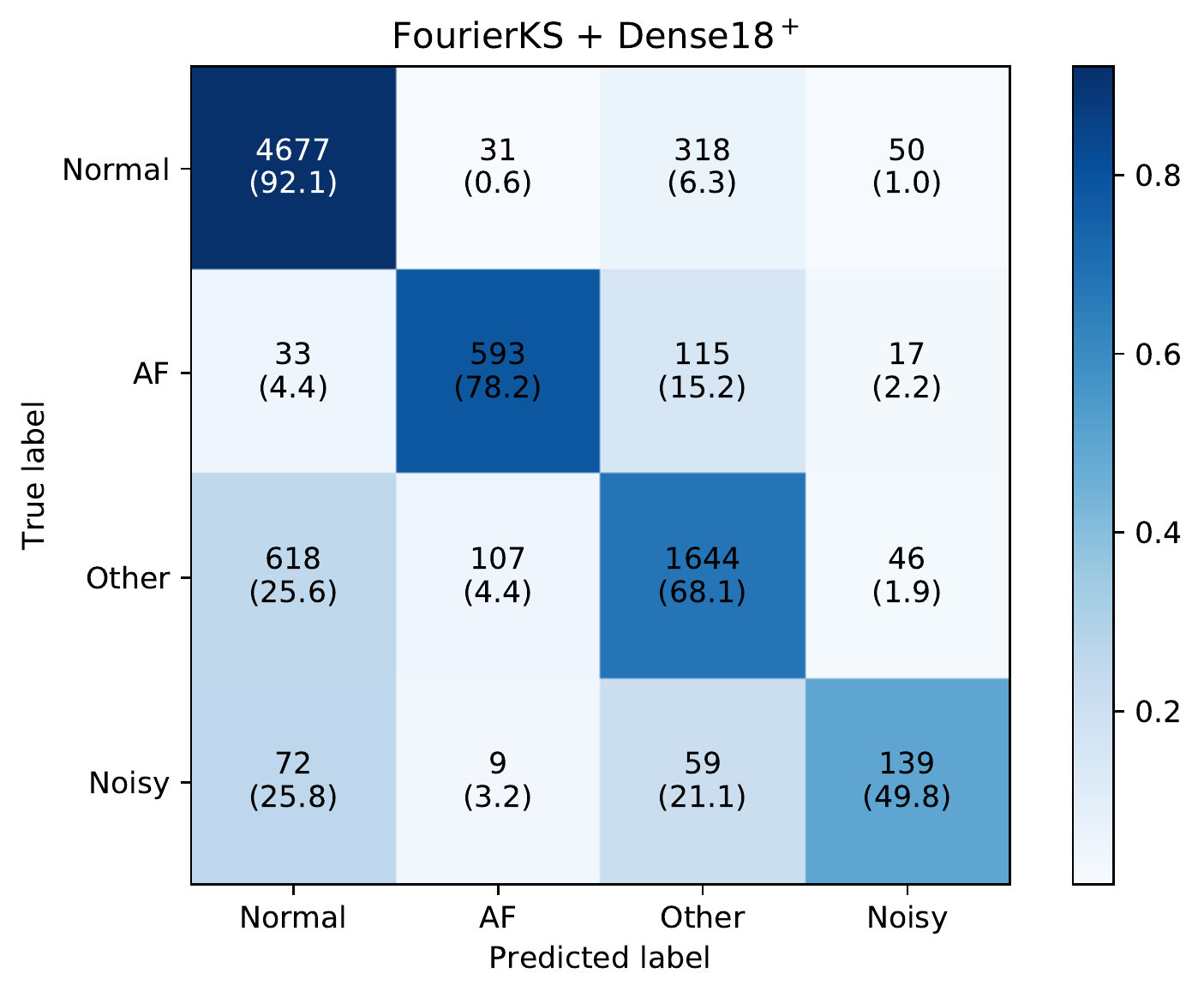}}
	\subfigure{\includegraphics[width=0.3\linewidth]{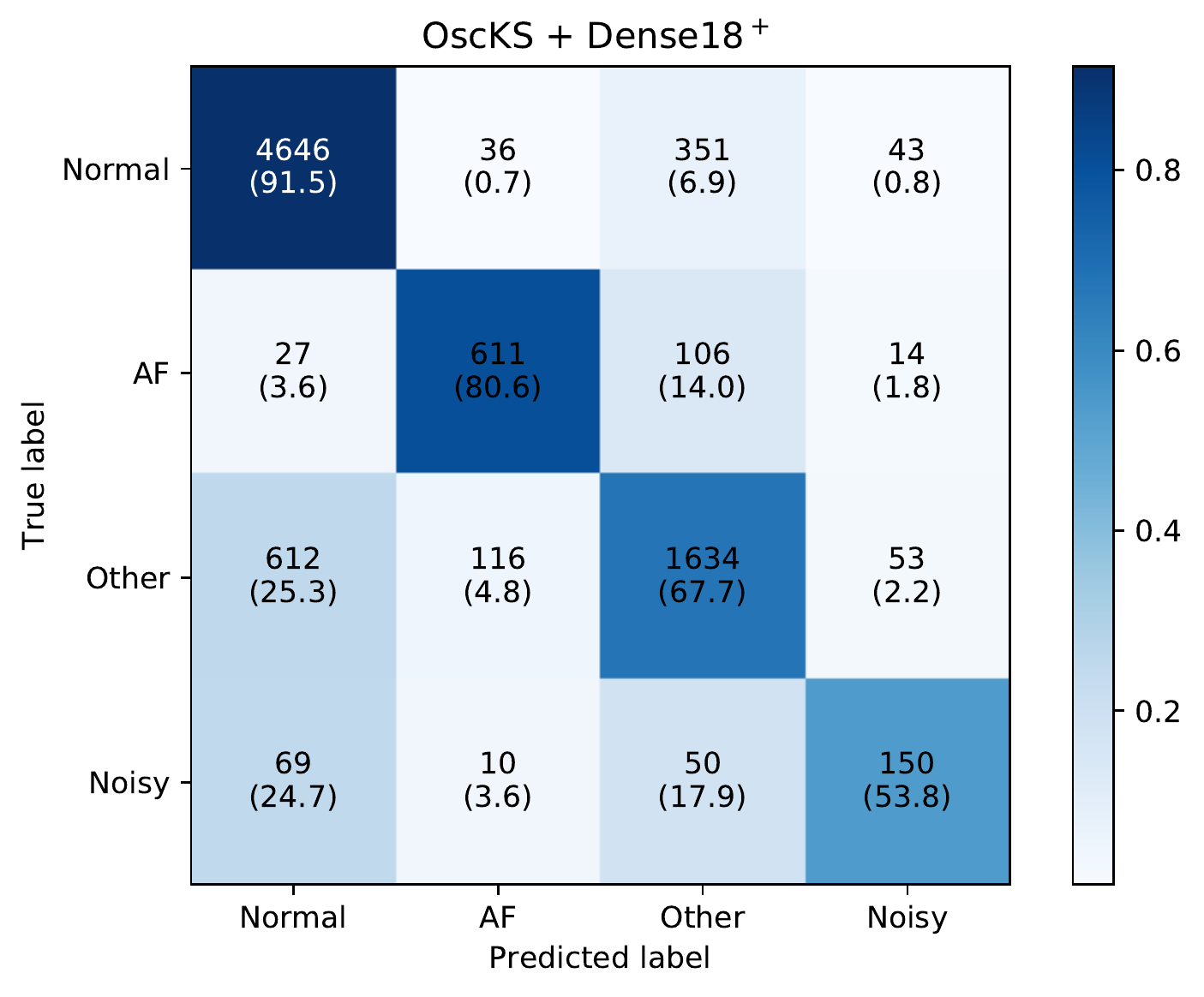}}
	\subfigure{\includegraphics[width=0.3\linewidth]{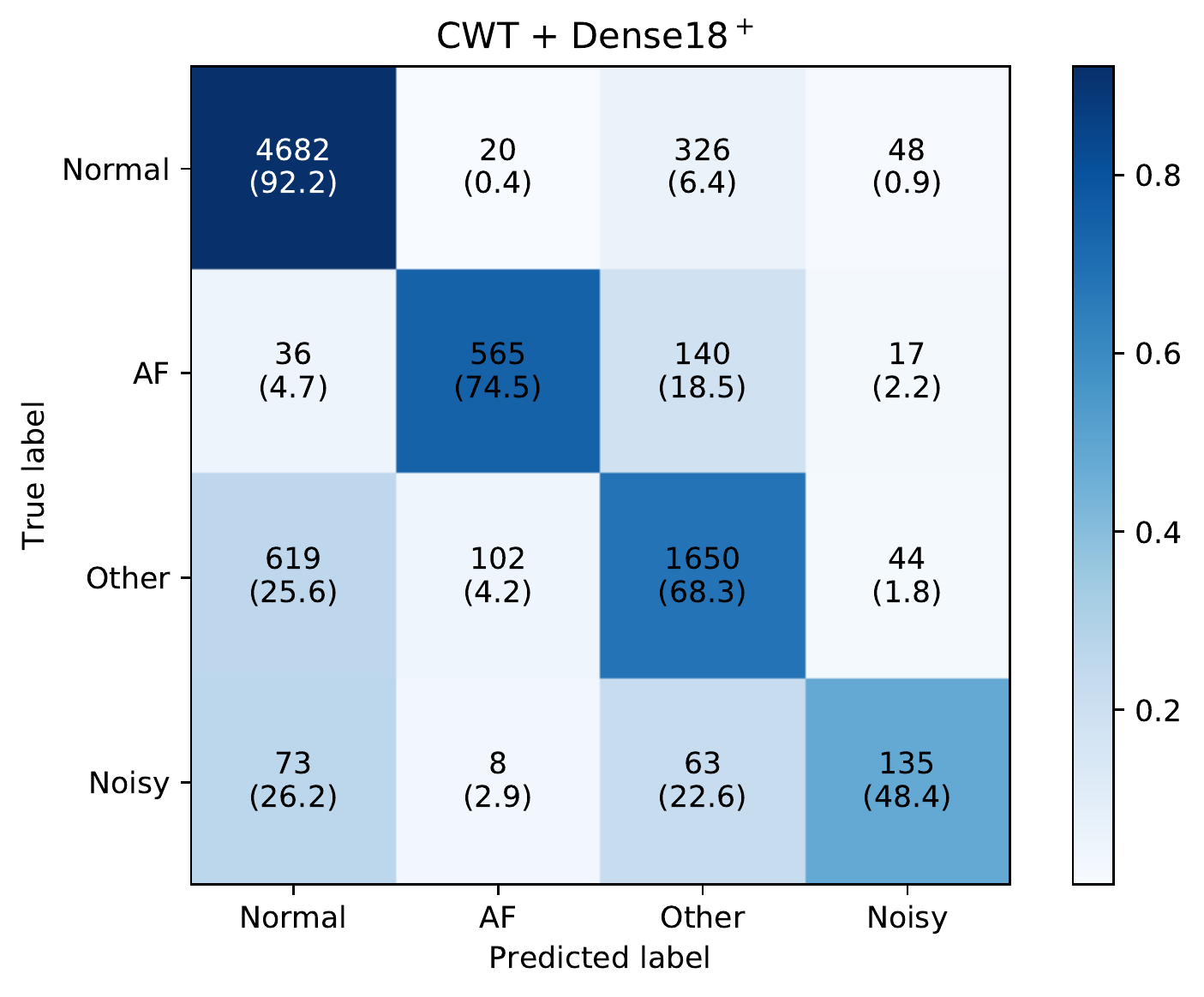}}
	\subfigure{\includegraphics[width=0.3\linewidth]{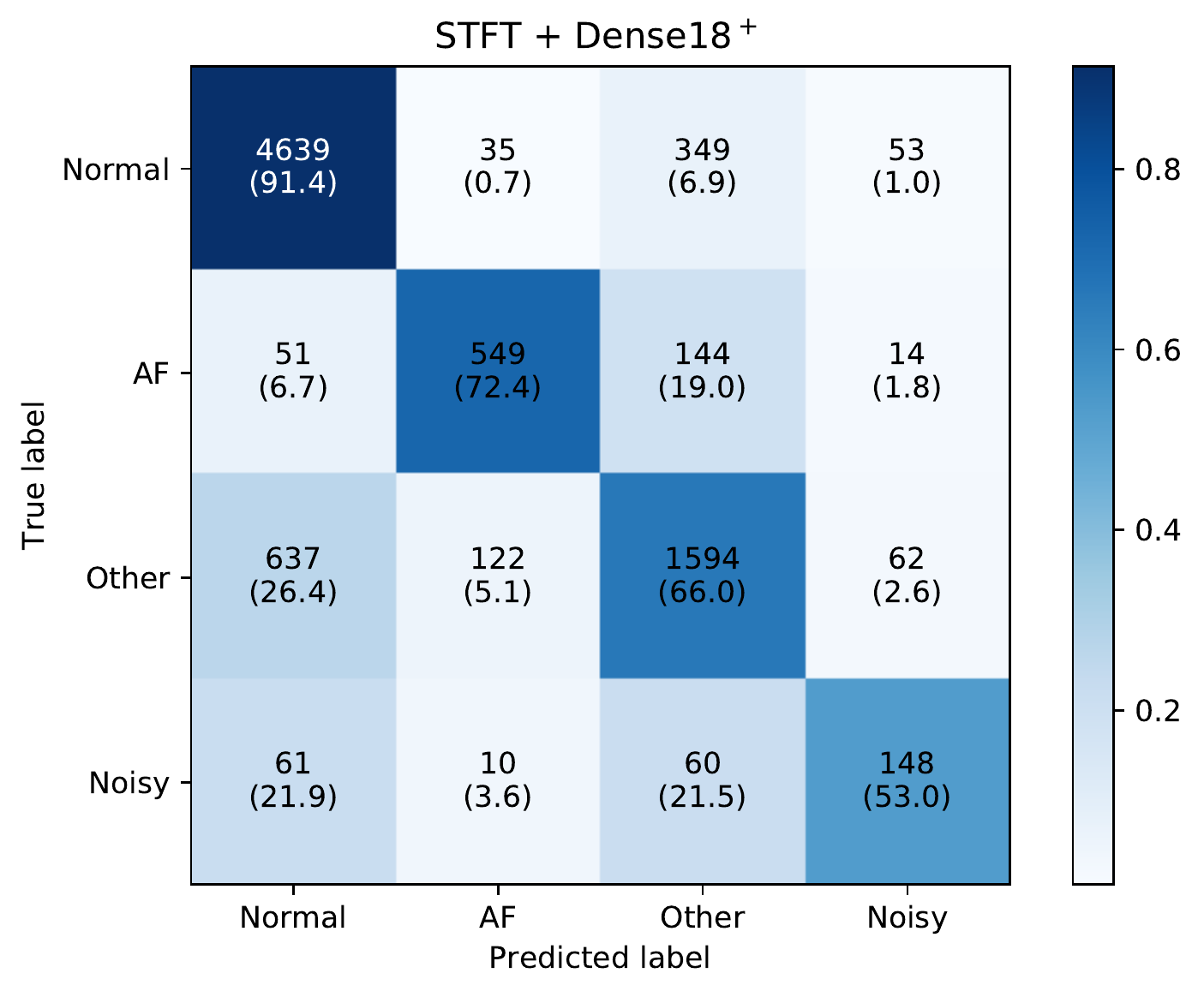}}
	\subfigure{\includegraphics[width=0.3\linewidth]{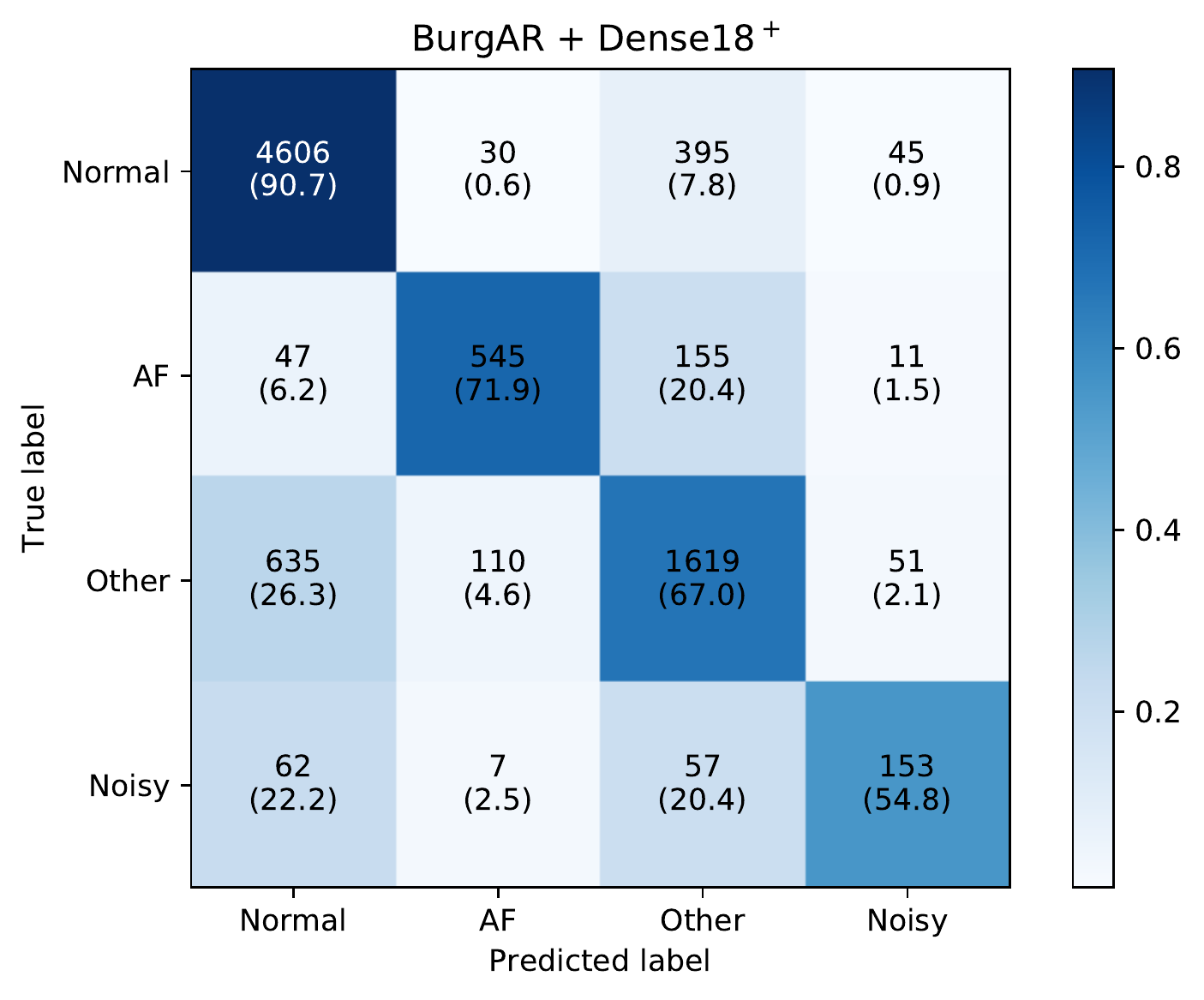}}
	\caption{Normalized confusion matrix on different methods.}
	\label{fig:cm}
\end{figure*}
\begin{figure*}[htb!]%
	\centering
	\subfigure[Atrial Fibrillation. Rec. 3223]{\label{fig:3223}\includegraphics[width=0.2\linewidth]{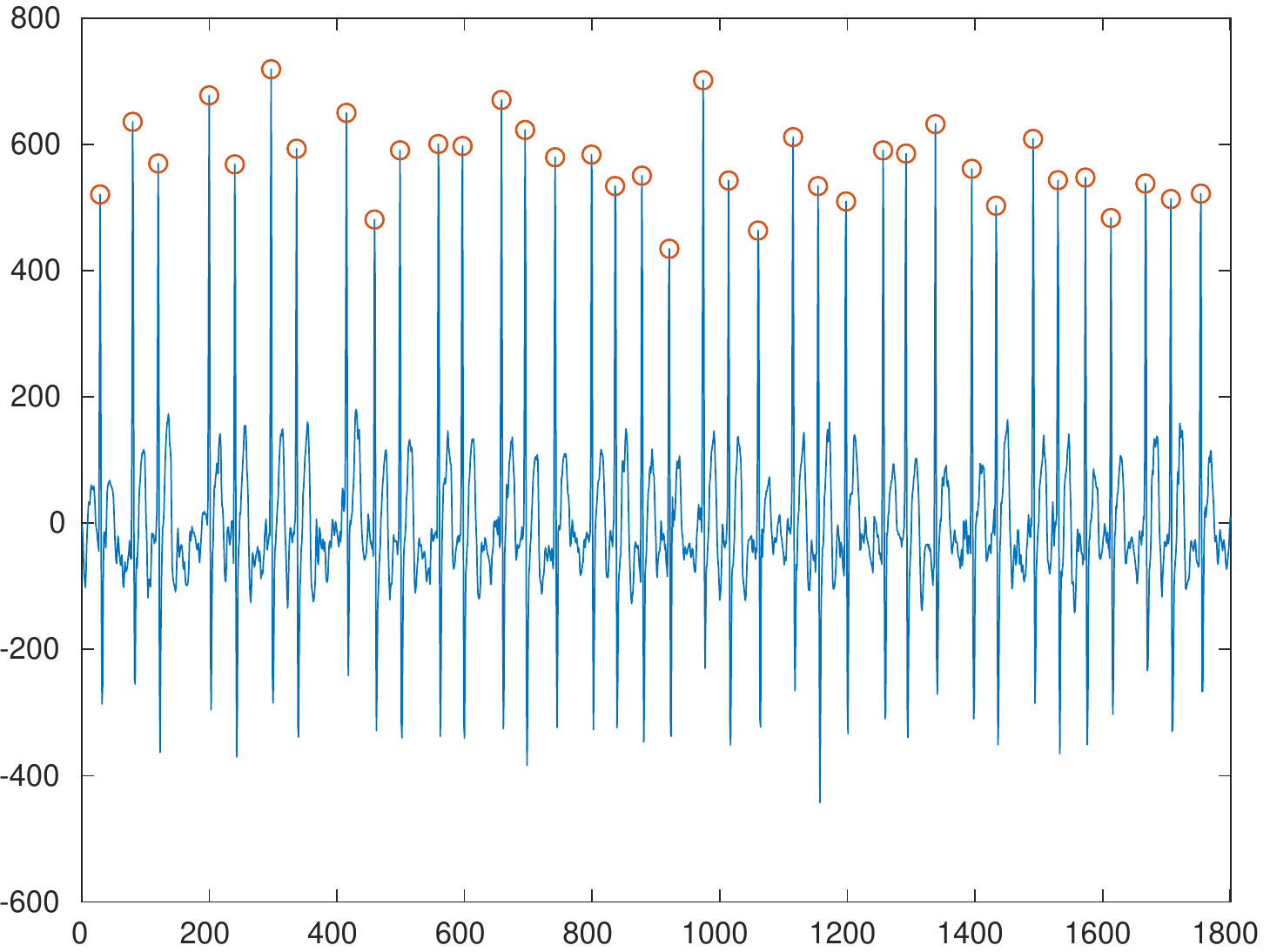}}
	\subfigure[FourierKS $\lambda = 10$; $M_1 = 0$, $M_{200} = 20$]{\label{fig:kf-smooth}\includegraphics[width=0.2\linewidth]{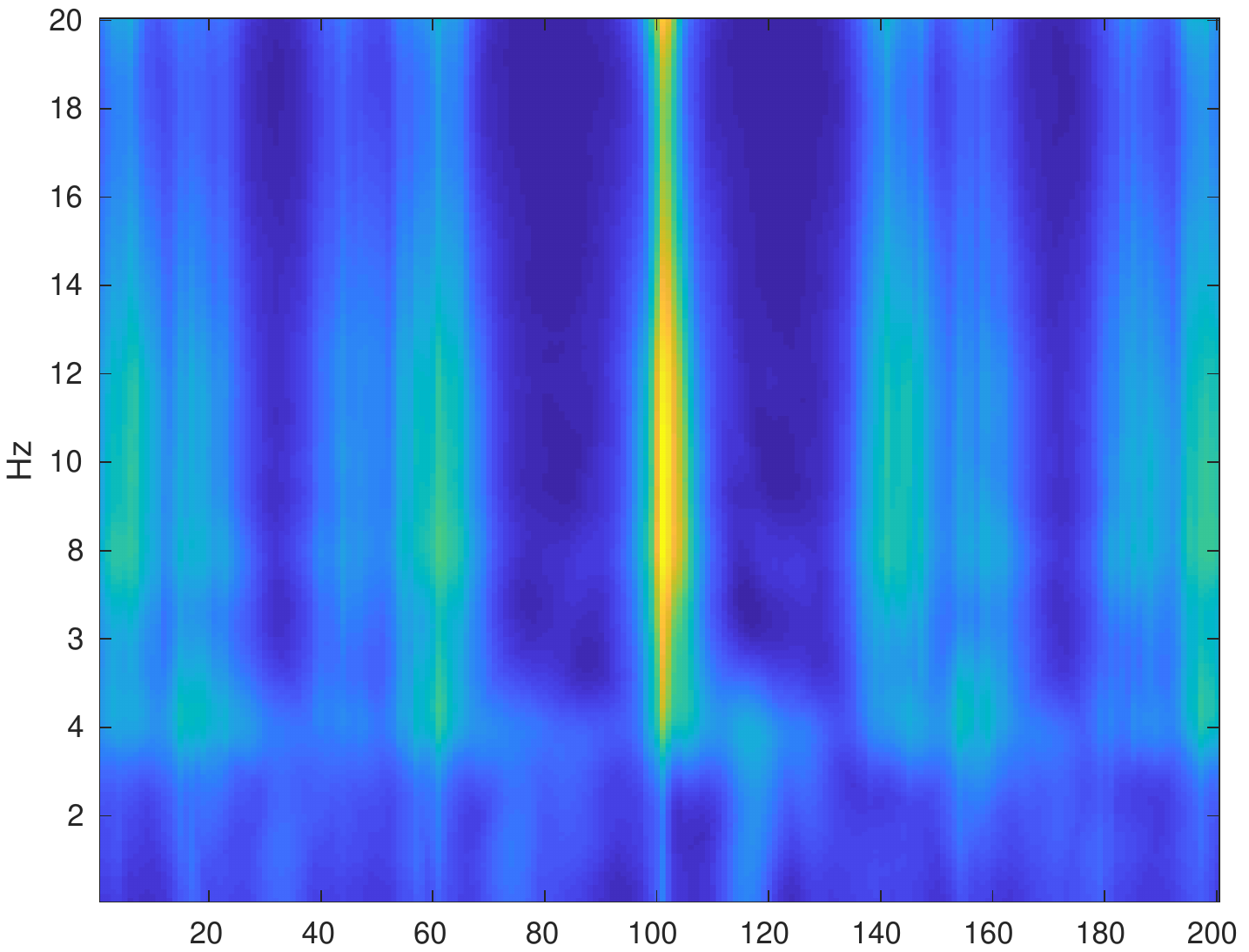}}
	\subfigure[FourierKS $\lambda = 10$; $M_1 = 0, M_{400} = 40$]{\label{fig:kalman-3246-smooth-large}\includegraphics[width=0.208\linewidth]{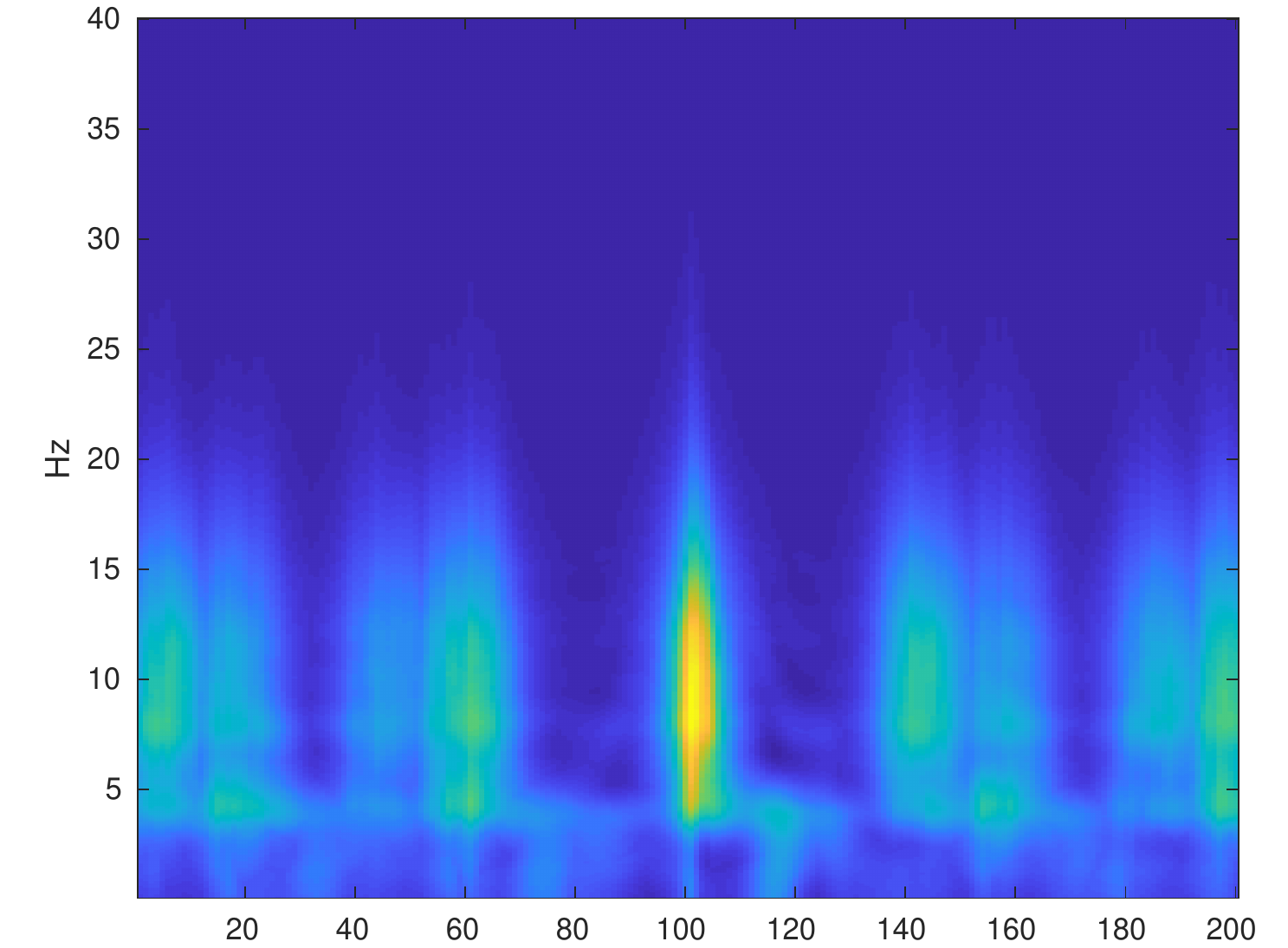}}
	\subfigure[OscKS $\lambda = 10, R=q=1, qb=1e^{-7}$; $M_1 = 0, M_{400} = 40$]{\label{fig:osc-3246-smooth}\includegraphics[width=0.2\linewidth]{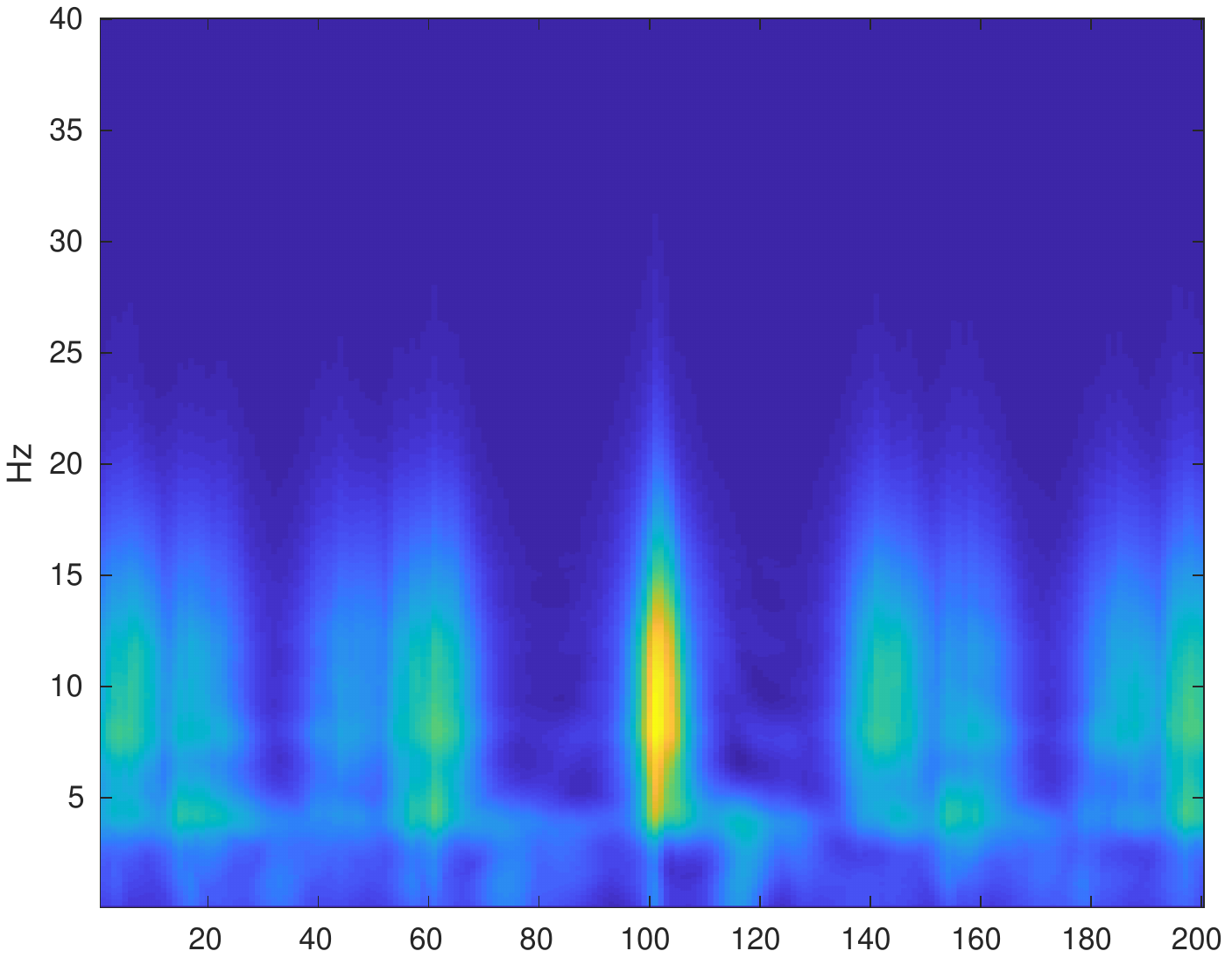}}
	\subfigure[Proposal in \cite{qi2002bayesian} $M_1 = 0, M_{400} = 40$ ]{\label{fig:kalman-3246-original}\includegraphics[width=0.203\linewidth]{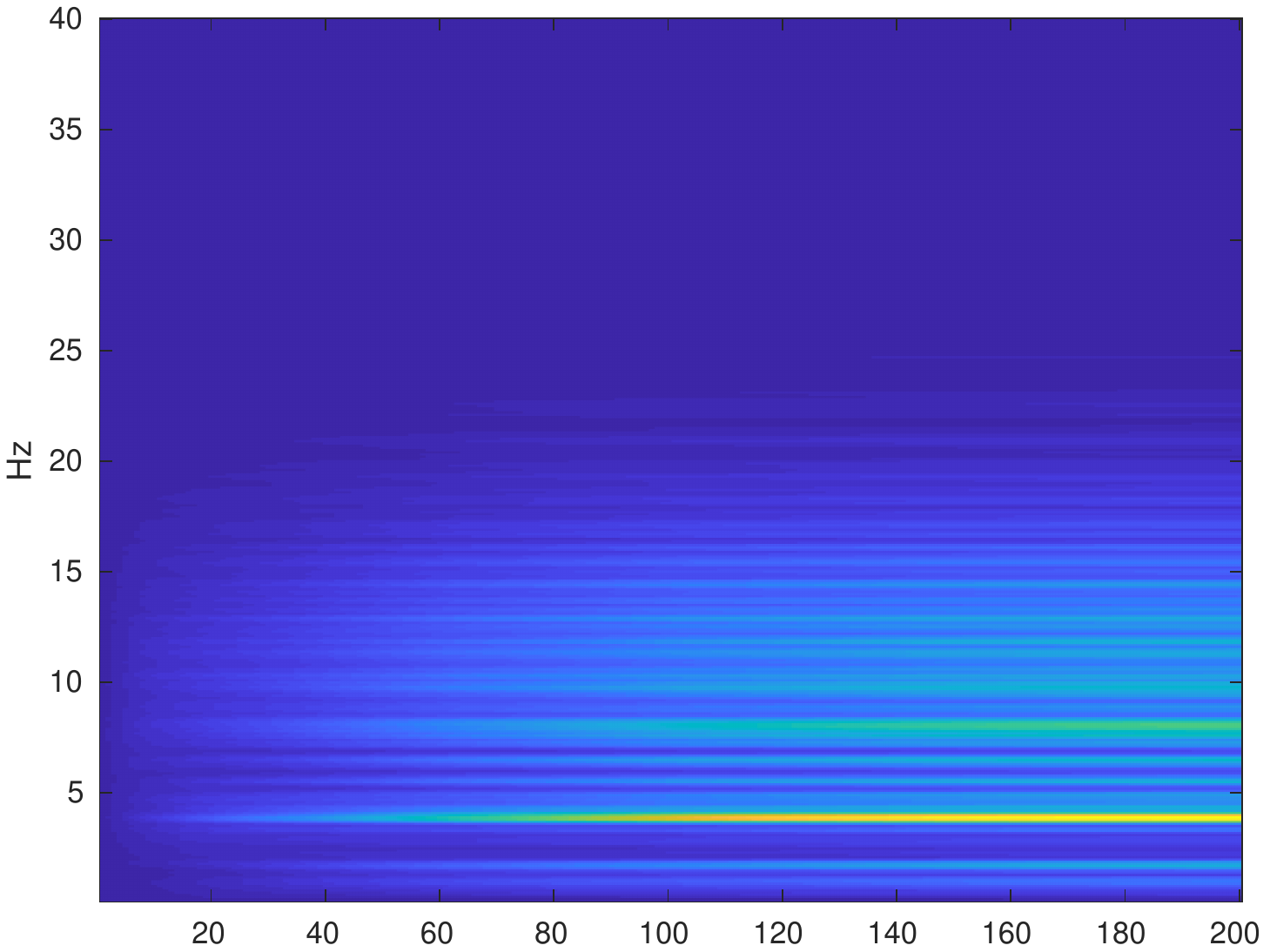}}
	\subfigure[STFT, Hann, window 11, overlap  10]{\label{fig:stft-3246}\includegraphics[width=0.2\linewidth]{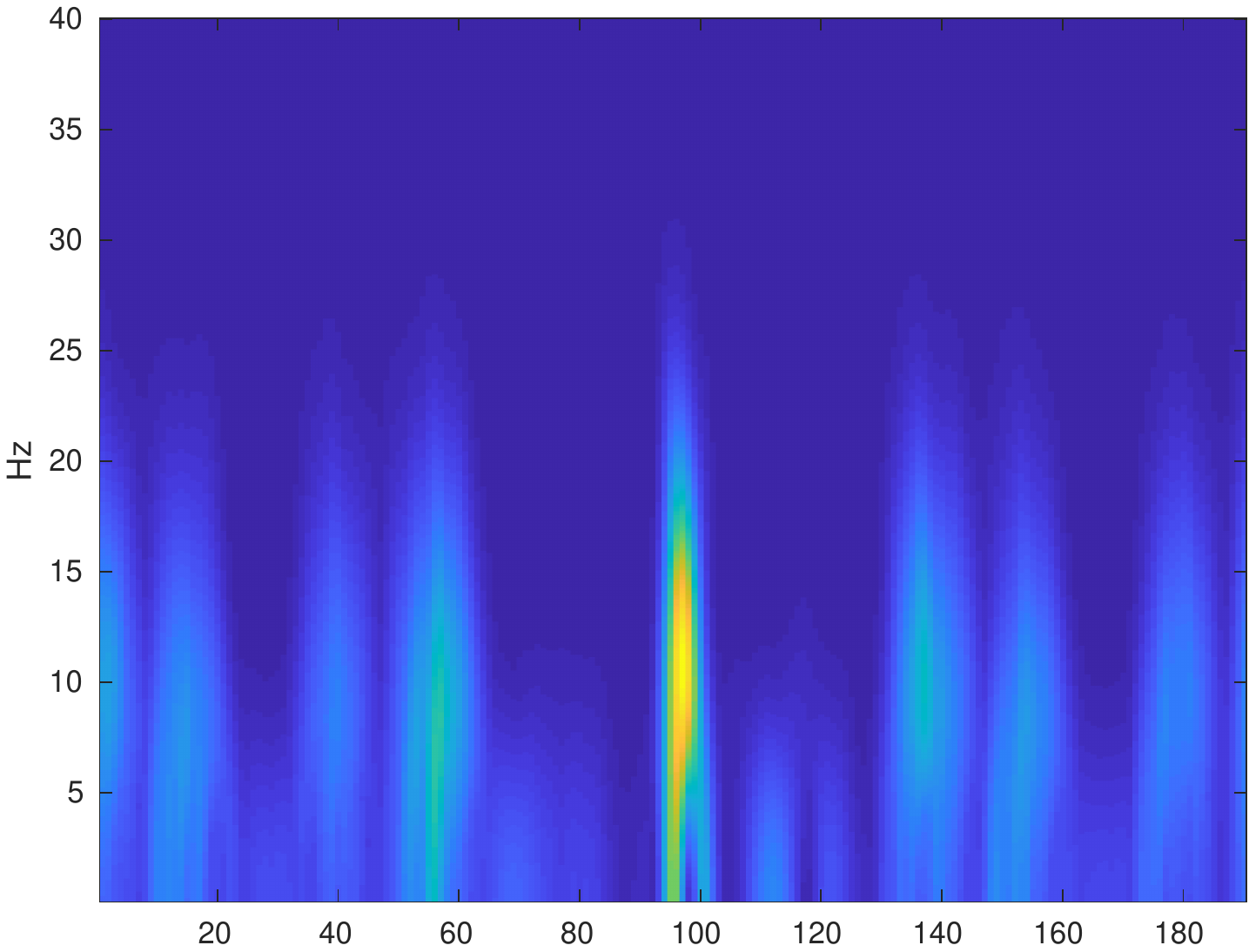}}
	\subfigure[CWT, wavelet morse]{\label{fig:cwt-3246}\includegraphics[width=0.2\linewidth]{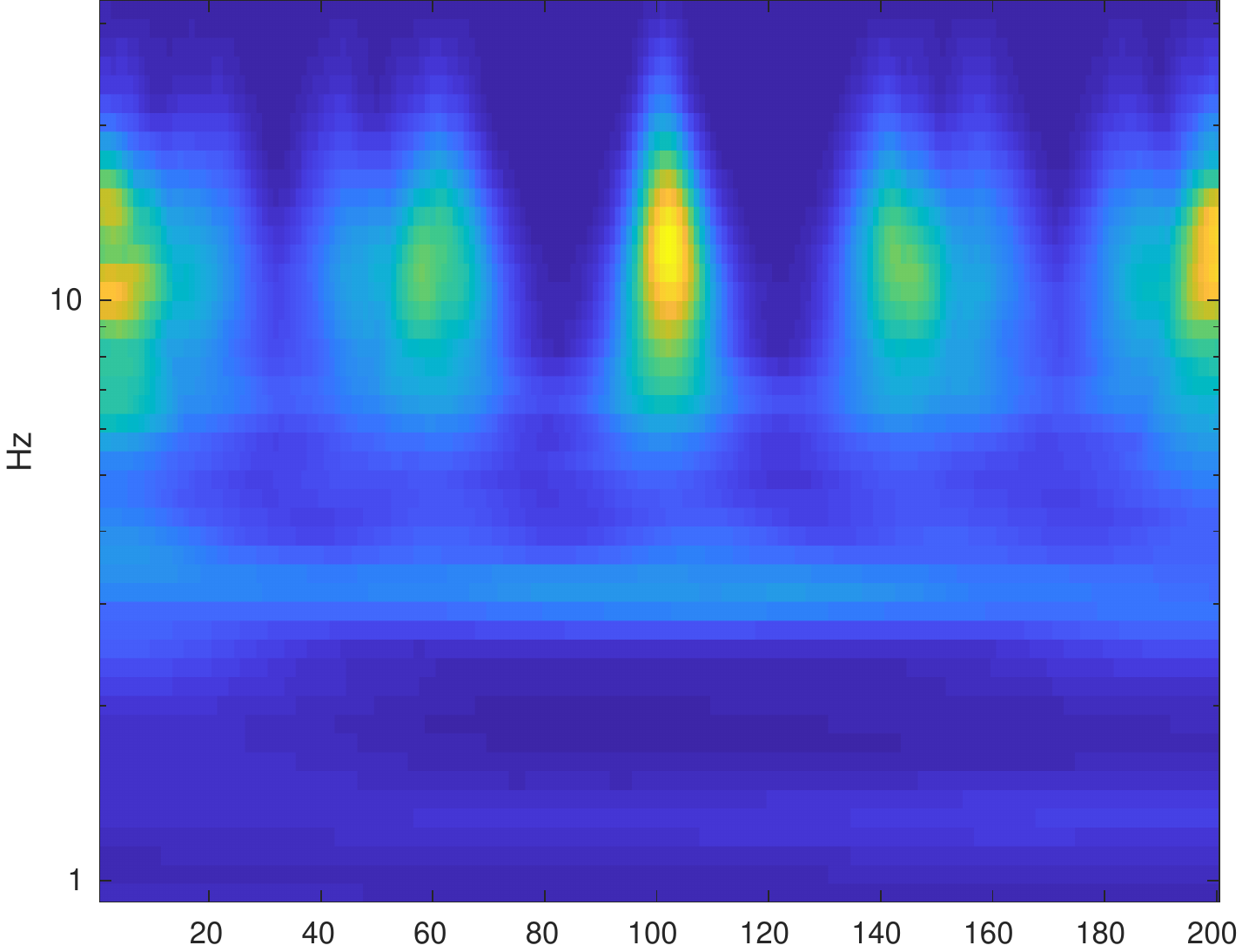}}
	\subfigure[BurgAR, Hann, window 11, overlap  10]{\label{fig:pburg-3246}\includegraphics[width=0.2\linewidth]{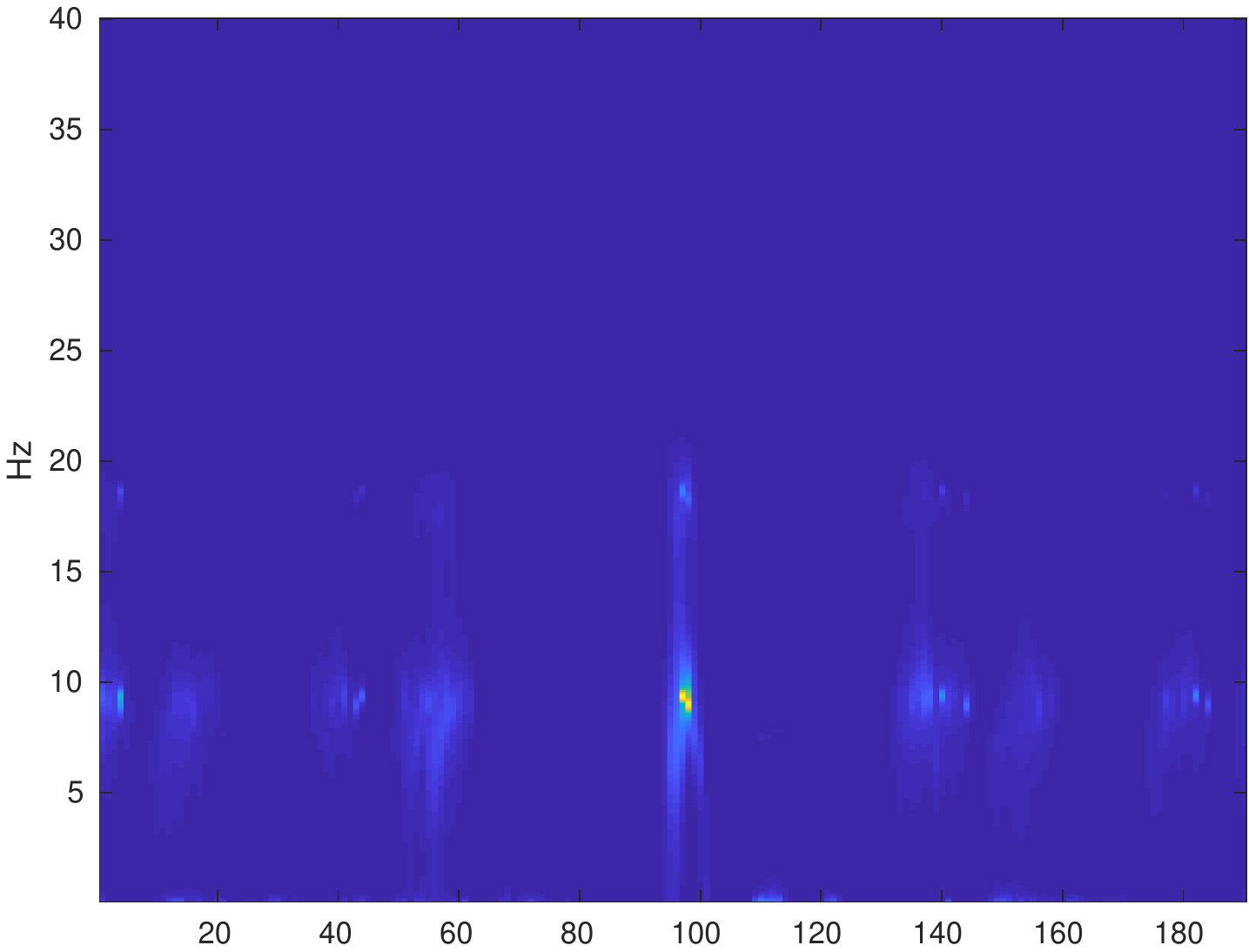}}
	\caption{Comparison of different spectrogram estimation methods on Rec. 3223. }
	\label{fig:kalman-spectrogram}
\end{figure*}

The detailed performance of all five methods 
(i.e., FourierKS, OscKS, CWT, STFT, and BurgAR) 
with Dense18$^+$ classifier are reported in five confusion matrices in Fig.~\ref{fig:cm}. Each confusion matrix is row-wise normalized. The diagonal entries show the Recall of each rhythm and off-diagonal entries show the misclassification rates. For example, the first row of the first confusion matrix shows 92.1\% of normal rhythms are correctly classified as normal, but 0.6\%, 6.3\%, and 1.0\% are incorrectly classified as AF, Other, and Noisy.

\section{Discussion}
\subsection{ECG Time-Frequency Analysis Methods}
\label{ssec:time-freq}

We first examine how different spectro-temporal estimation methods perform on an ECG signal through a visual inspection. We take the 3223th recording (Rec. 3223) from CinC 2017 dataset as example, which is labelled as AF. It is shown in Fig.\ref{fig:3223}. For the FourierKS and OscKS method, we choose different frequency range ($M$) and smoothing option as shown in Fig.~\ref{fig:kf-smooth}, \ref{fig:kalman-3246-smooth-large} and \ref{fig:osc-3246-smooth}. We set the length scale $\lambda$ to a constant 10, and use 1 for variance of measurement noise $R$, and identity for covariance of process noise $q$. In theory, $\lambda$ could be different for each frequency, which could be used to improve the performance.  Fig.~\ref{fig:kalman-3246-original} presents results by the original method in \cite{qi2002bayesian}, which adopts Brownian motion model for the coefficients. For STFT and BurgAR, we apply apply 11 length 10 overlapping Hann windows for estimation, as shown in Fig.~\ref{fig:stft-3246} and \ref{fig:pburg-3246}. For CWT (Fig.~\ref{fig:cwt-3246}), we use the default Morse wavelet implemented in Matlab. 

First, we observe that the estimation results of FourierKS (Fig.\ref{fig:kalman-3246-smooth-large}) and OscKS (Fig.\ref{fig:osc-3246-smooth}) are nearly the same except that the base frequency $a_0$ coefficient estimates are very sensitive to $q_b$ in the OscKS method. If we compare FourierKS method to STFT, BurgAR, and CWT, which are shown in Fig.~\ref{fig:kalman-3246-smooth-large}, \ref{fig:stft-3246}, \ref{fig:pburg-3246}, and \ref{fig:cwt-3246} respectively, we can initially conclude several advantages: the result from FourierKS is more smooth and it has higher and more unified resolution on both time and frequency. For STFT and BurgAR, the resolution is confined by window selection, length, and overlap. CWT untangles this problem by scaling and translation of wavelet basis function, but due to uncertainty principle of wavelet signal processing \cite{Ricaud2014}, the required resolution in time and frequency can not be met simultaneously (see Fig.\ref{fig:cwt-3246}). Our approaches model the time-varying Fourier series coefficients of signal in state-space, which are free from usage of windows or wavelets. 

Another advantage of the proposed OscKS estimation method is that it can be very computationally efficient for implementation when we need to perform estimation many times and the system is fixed (i.e., $\cu{A}$, $\cu{Q}$ remain unchanged). For example, if one takes the averaging strategy, the spectrum estimation has to be done for every segment and recording. For OscKS method, we merely need to solve $\cu{P}_\infty$ in \eqref{equ:riccati-rc} once. As we stated in Section \ref{sec:osc}, the computational cost of OscKS method is substantial reduced by deriving a stable covariance.

\subsection{ECG Classification for AF Detection}\label{MethodComparison}

As it is mentioned before, Table~\ref{tbl:f1-overall} shows that the best results belong  to  our  proposed spectro-temporal  representation  methods  (i.e.,  FourierKS and OscKS) with Dense18$^+$ classifier. Table~\ref{tbl:f1-overall} also shows that independent of spectro-temporal representation method, Dense18$^+$ has the highest performance among all classifiers. In contrast, the plain CNN (CNN18) has the lowest scores. In addition, RF is generally worse than convolutional networks classifiers (except CNN18) probably because in contrary to convolutional networks, RF has not benefited from the existing structure in spectro-temporal representation. 	

Regarding the different spectro-temporal representations STFT and BurgAR have the worst results, and FourierKS, and OscKS have the best performance. In addition, for some classifiers CWT provides the results which are as good as or even better than FourierKS, and OscKS. However, the best results of FourierKS, and OscKS are higher than the best result of CWT.

Table~\ref{tbl:detail-compare} shows that the the proposed ECG classification methods have the best result for Normal rhythm and the worst result for Noisy. The performance of AF and Other are between these two, but typically AF has better performance that Other, probably because Other is an umberella term that covers many abnormal non-AF rhythms, and we do not have enough samples for each abnormalities to properly train our classifiers.

\begin{figure*}[h!]
	\centering
	\includegraphics[scale=0.4]{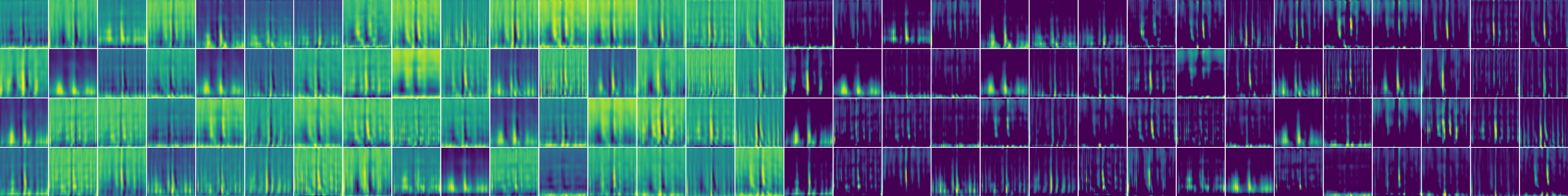}
	\includegraphics[scale=0.4]{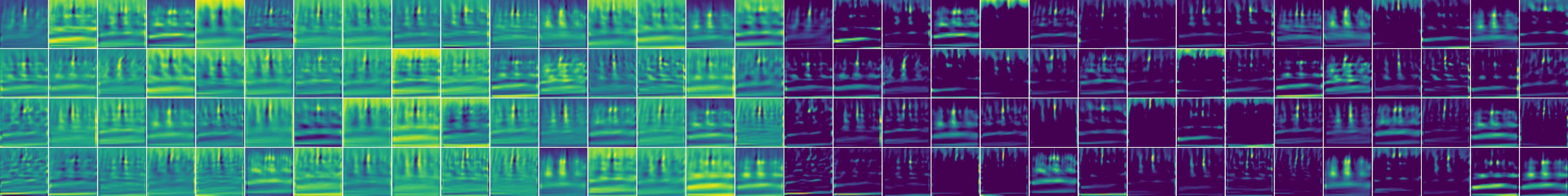}
	\includegraphics[scale=0.4]{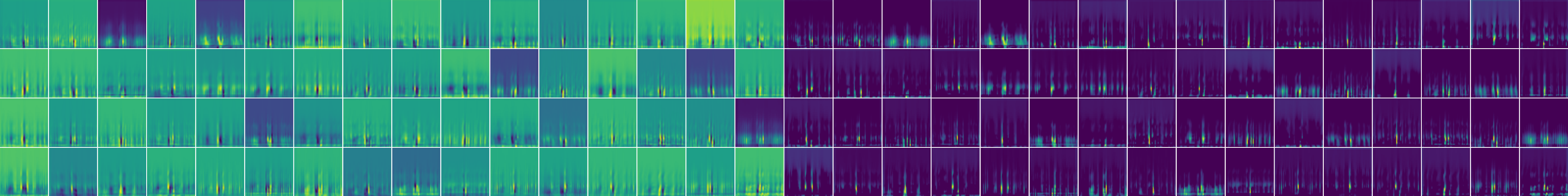}
	\includegraphics[scale=0.4]{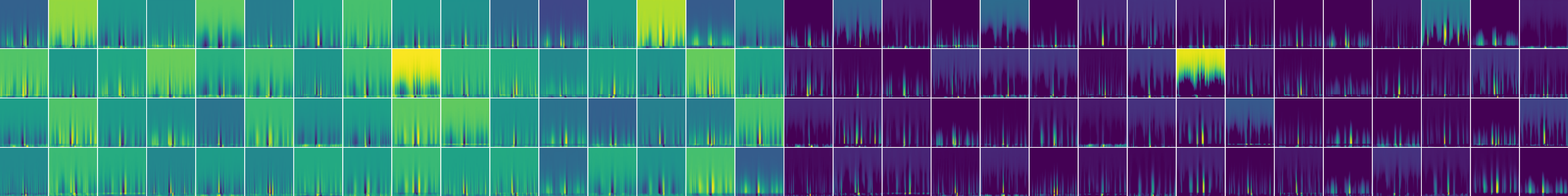}
	\caption{Feature-map (Left 16 columns) and activation (right 16 columns) visualization of first convolutional layer on Rec.~1005 (AF). From top to bottom, every 4 rows are FourierKS, CWT, BurgAR and STFT respectively. OscKS is not shown here for simplicity, because it has a very similar result to FourierKS.}
	\label{fig:feature_map}
\end{figure*}

To examine how different spectro-temporal features act in AF ECG analysis, one elementary-level way is to investigate the feature map and activation of the first convolutional layer. However, this voxel-based ``probing'' only produces limited explanation \cite{Szegedy14intriguingproperties}, and can not fully give the insights. The visualization is shown in Fig.~\ref{fig:feature_map}. We can see that the feature-map of FourierKS and CWT are more diverse and active than STFT and BurgAR, and they have larger activation on ``peaks'' and background details. In comparison to FourierKS and CWT, the lower-frequency area are better preserved and exploited for FourierKS method.

\subsection{Limitations}
\label{sec:limitation}

Typically for AF detection we need at least 30~s ECG data~\cite{developed2010guidelines}. However, many ECG recordings in the dataset have less than 30~s duration (see Section~\ref{new:dataset}) which limits the medical significance of the current study. In addition, the averaging step in feature engineering is robust only when there are enough spectro-temporal segments, which is not the case for very short ECG recordings (see Section~\ref{ssec:ave-overall}).

\section{Conclusion}
\label{sec:majhead}

In this paper, we proposed a spectro-temporal representation of ECG signals, based on state-space models, for application in deep network based atrial fibrillation detection. We empirically showed that if we put Gaussian process priors on the Fourier series coeffients, then by estimating the state of the corresponding linear state-space model using Kalman filter/smoother we can outperform other time-frequency analysis methods such as short-time Fourier transform, continuous wavelet transform, and autoregressive spectral estimation for ECG classification.

We also accelerated the estimation of the spectro-temporal representation of signals by using a stochastic oscillator differential equation model and stationary Kalman filter/smoother. This representation is useful to improve the scalability of the proposed spectro-temporal representation for long ECG recordings. Finally, we have found an efficient convolutional architecture (i.e., Dense18$^+$) for AF detection using the spectro-temporal features by comparative evaluation of multiple convolutional neural networks models.

\bibliographystyle{spmpsci}      
\bibliography{refs.bib}  

\begin{thebibliography}{10}
\providecommand{\url}[1]{{#1}}
\providecommand{\urlprefix}{URL }
\expandafter\ifx\csname urlstyle\endcsname\relax
  \providecommand{\doi}[1]{DOI~\discretionary{}{}{}#1}\else
  \providecommand{\doi}{DOI~\discretionary{}{}{}\begingroup
  \urlstyle{rm}\Url}\fi

\bibitem{ANNAVARAPU2016151}
Annavarapu, A., Kora, P.: {ECG}-based atrial fibrillation detection using
  different orderings of conjugate symmetric–complex {H}adamard transform.
\newblock International Journal of the Cardiovascular Academy \textbf{2}(3),
  151--154 (2016)

\bibitem{asgari2015automatic}
Asgari, S., Mehrnia, A., Moussavi, M.: Automatic detection of atrial
  fibrillation using stationary wavelet transform and support vector machine.
\newblock Computers in Biology and Medicine \textbf{60}, 132--142 (2015)

\bibitem{BABAEIZADEH2009522}
Babaeizadeh, S., Gregg, R.E., Helfenbein, E.D., Lindauer, J.M., Zhou, S.H.:
  Improvements in atrial fibrillation detection for real-time monitoring.
\newblock Journal of Electrocardiology \textbf{42}(6), 522--526 (2009)

\bibitem{liaw2002classification}
Breiman, L.: Random forests.
\newblock Machine Learning \textbf{45}(1), 5--32 (2001)

\bibitem{bruser2013automatic}
Bruser, C., Diesel, J., Zink, M.D., Winter, S., Schauerte, P., Leonhardt, S.:
  Automatic detection of atrial fibrillation in cardiac vibration signals.
\newblock IEEE Journal of Biomedical and Health Informatics \textbf{17}(1),
  162--171 (2013)

\bibitem{developed2010guidelines}
Camm, A.J., Kirchhof, P., Lip, G.Y., Schotten, U., Savelieva, I., Ernst, S.,
  Van~Gelder, I.C., Al-Attar, N., Hindricks, G., Prendergast, B., et~al.:
  Guidelines for the management of atrial fibrillation: the task force for the
  management of atrial fibrillation of the european society of cardiology
  {(ESC)}.
\newblock European Heart Journal \textbf{31}(19), 2369--2429 (2010)

\bibitem{clifford2017af}
Clifford, G.D., et~al.: {AF} classification from a short single lead {ECG}
  recording: the {P}hysionet/{C}omputing in {C}ardiology {C}hallenge 2017.
\newblock 2017 Computing in Cardiology (CinC) \textbf{44}, 1--4 (2017)

\bibitem{ehrendorfer2012spectral}
Ehrendorfer, M.: Spectral Numerical Weather Prediction Models.
\newblock Society for Industrial and Applied Mathematics (2011)

\bibitem{GARCIA2016157}
García, M., Ródenas, J., Alcaraz, R., Rieta, J.J.: Application of the
  relative wavelet energy to heart rate independent detection of atrial
  fibrillation.
\newblock Computer Methods and Programs in Biomedicine \textbf{131}, 157--168
  (2016)

\bibitem{glorot2010understanding}
Glorot, X., Bengio, Y.: Understanding the difficulty of training deep
  feedforward neural networks.
\newblock In: Proceedings of the 13th International Conference on Artificial
  Intelligence and Statistics, vol.~9, pp. 249--256 (2010)

\bibitem{goodfellow2016deep}
Goodfellow, I., Bengio, Y., Courville, A., Bengio, Y.: Deep Learning.
\newblock MIT Press (2016)

\bibitem{Grewal+Andrews:2001}
Grewal, M.S., Andrews, A.P.: Kalman Filtering, Theory and Practice Using
  {MATLAB}.
\newblock Wiley, New York, NY (2001)

\bibitem{HAGIWARA201899}
Hagiwara, Y., Fujita, H., Oh, S.L., Tan, J.H., Tan, R.S., Ciaccio, E.J.,
  Acharya, U.R.: Computer-aided diagnosis of atrial fibrillation based on {ECG}
  signals: {A} review.
\newblock Information Sciences \textbf{467}, 99--114 (2018)

\bibitem{Hartikainen+Sarkka:2010}
Hartikainen, J., S\"arkk\"a, S.: Kalman filtering and smoothing solutions to
  temporal {G}aussian process regression models.
\newblock In: 2010 IEEE International Workshop on Machine Learning for Signal
  Processing (MLSP), pp. 379--384 (2010)

\bibitem{he2016deep}
He, K., Zhang, X., Ren, S., Sun, J.: Deep residual learning for image
  recognition.
\newblock In: 2016 IEEE Conference on Computer Vision and Pattern Recognition
  (CVPR), pp. 770--778 (2016)

\bibitem{huang2017densely}
Huang, G., Liu, Z., van~der Maaten, L., Weinberger, K.Q.: Densely connected
  convolutional networks.
\newblock In: 2017 IEEE Conference on Computer Vision and Pattern Recognition
  (CVPR), pp. 2261--2269 (2017)

\bibitem{joseph2017daily}
Joseph, A., Larrain, M., Turner, C.: Daily stock returns characteristics and
  forecastability.
\newblock Procedia Computer Science \textbf{114}, 481--490 (2017)

\bibitem{Kailath:233814}
Kailath, T., Sayed, A.H., Hassibi, B.: Linear Estimation.
\newblock Prentice Hall, New Jersey (2000)

\bibitem{kay1981spectrum}
Kay, S.M., Marple, S.L.: Spectrum analysis -- a modern perspective.
\newblock Proceedings of the IEEE \textbf{69}(11), 1380--1419 (1981)

\bibitem{NIPS2012_4824}
Krizhevsky, A., Sutskever, I., Hinton, G.E.: {ImageNet} classification with
  deep convolutional neural networks.
\newblock In: Advances in Neural Information Processing Systems 25, pp.
  1097--1105. Curran Associates, Inc. (2012)

\bibitem{lecun2015deep}
LeCun, Y., Bengio, Y., Hinton, G.: Deep learning.
\newblock Nature \textbf{521}(7553), 436--444 (2015)

\bibitem{mohebbi2008detection}
Mohebbi, M., Ghassemian, H.: Detection of atrial fibrillation episodes using
  {SVM}.
\newblock In: 2008 30th Annual International Conference of the IEEE Engineering
  in Medicine and Biology Society, pp. 177--180. IEEE (2008)

\bibitem{pan1985real}
Pan, J., Tompkins, W.J.: A real-time {QRS} detection algorithm.
\newblock IEEE Transactions on Biomedical Engineering \textbf{BME-32}(3),
  230--236 (1985)

\bibitem{pourbabaee2017deep}
Pourbabaee, B., Roshtkhari, M.J., Khorasani, K.: Deep convolutional neural
  networks and learning {ECG} features for screening paroxysmal atrial
  fibrillation patients.
\newblock IEEE Transactions on Systems, Man, and Cybernetics: Systems
  \textbf{48}(12), 2095--2104 (2018)

\bibitem{qi2002bayesian}
Qi, Y., Minka, T.P., Picara, R.W.: Bayesian spectrum estimation of unevenly
  sampled nonstationary data.
\newblock In: 2002 IEEE International Conference on Acoustics, Speech, and
  Signal Processing (ICASSP), vol.~2, pp. 1473--1476. IEEE (2002)

\bibitem{rad2012phase}
Rad, A.B., Virtanen, T.: Phase spectrum prediction of audio signals.
\newblock In: 2012 5th International Symposium on Communications, Control and
  Signal Processing, pp. 1--5. IEEE (2012)

\bibitem{rad2017ecg}
Rad, A.B., et~al.: {ECG}-based classification of resuscitation cardiac rhythms
  for retrospective data analysis.
\newblock IEEE Transactions on Biomedical Engineering \textbf{64}(10),
  2411--2418 (2017)

\bibitem{rajpurkar2017cardiologist}
Rajpurkar, P., Hannun, A.Y., Haghpanahi, M., Bourn, C., Ng, A.Y.:
  Cardiologist-level arrhythmia detection with convolutional neural networks.
\newblock arXiv preprint arXiv:1707.01836  (2017)

\bibitem{Ricaud2014}
Ricaud, B., Torr{\'e}sani, B.: A survey of uncertainty principles and some
  signal processing applications.
\newblock Advances in Computational Mathematics \textbf{40}(3), 629--650 (2014)

\bibitem{8331569}
Rubin, J., Parvaneh, S., Rahman, A., Conroy, B., Babaeizadeh, S.: Densely
  connected convolutional networks and signal quality analysis to detect atrial
  fibrillation using short single-lead {ECG} recordings.
\newblock In: 2017 Computing in Cardiology (CinC), pp. 1--4 (2017)

\bibitem{sarkka2013bayesian}
S{\"a}rkk{\"a}, S.: Bayesian Filtering and Smoothing.
\newblock Cambridge University Press (2013)

\bibitem{Sarkka+Solin+Hartikainen:2013}
S\"arkk\"a, S., Solin, A., Hartikainen, J.: Spatiotemporal learning via
  infinite-dimensional {B}ayesian fltering and smoothing.
\newblock IEEE Signal Processing Magazine \textbf{30}(4), 51--61 (2013)

\bibitem{SARKKA20121517}
S{\"a}rkk{\"a}, S., Solin, A., Nummenmaa, A., Vehtari, A., Auranen, T., Vanni,
  S., Lin, F.H.: Dynamic retrospective filtering of physiological noise in
  {BOLD} {fMRI}: {DRIFTER}.
\newblock NeuroImage \textbf{60}(2), 1517--1527 (2012)

\bibitem{shashikumar2017deep}
Shashikumar, S.P., Shah, A.J., Li, Q., Clifford, G.D., Nemati, S.: A deep
  learning approach to monitoring and detecting atrial fibrillation using
  wearable technology.
\newblock In: 2017 IEEE EMBS International Conference on Biomedical Health
  Informatics (BHI), pp. 141--144. IEEE (2017)

\bibitem{solin14}
Solin, A., Särkkä, S.: {Explicit Link Between Periodic Covariance Functions
  and State Space Models}.
\newblock In: S.~Kaski, J.~Corander (eds.) Proceedings of the Seventeenth
  International Conference on Artificial Intelligence and Statistics,
  \emph{Proceedings of Machine Learning Research}, vol.~33, pp. 904--912. PMLR,
  Reykjavik, Iceland (2014)

\bibitem{AAAI1714806}
Szegedy, C., Ioffe, S., Vanhoucke, V., Alemi, A.: Inception-v4,
  {Inception-ResNet} and the impact of residual connections on learning.
\newblock In: Proceedings of AAAI on Artificial Intelligence, pp. 4278--4284
  (2017)

\bibitem{Szegedy_2016_CVPR}
Szegedy, C., Vanhoucke, V., Ioffe, S., Shlens, J., Wojna, Z.: Rethinking the
  inception architecture for computer vision.
\newblock In: The IEEE Conference on Computer Vision and Pattern Recognition
  (CVPR) (2016)

\bibitem{Szegedy14intriguingproperties}
Szegedy, C., Zaremba, W., Sutskever, I., Bruna, J., Erhan, D., Goodfellow, I.,
  Fergus, R.: Intriguing properties of neural networks.
\newblock arXiv preprint arXiv:1312.6199  (2013)

\bibitem{thaler2017only}
Thaler, M.: The Only {EKG} Book You'll Ever Need.
\newblock Lippincott Williams \& Wilkins (2017)

\bibitem{xia2018detecting}
Xia, Y., Wulan, N., Wang, K., Zhang, H.: Detecting atrial fibrillation by deep
  convolutional neural networks.
\newblock Computers in Biology and Medicine \textbf{93}, 84--92 (2018)

\bibitem{xiong2017robust}
Xiong, Z., Stiles, M.K., Zhao, J.: Robust {ECG} signal classification for
  detection of atrial fibrillation using a novel neural network.
\newblock 2017 Computing in Cardiology (CinC) \textbf{44}, 1--4 (2017)

\bibitem{yaghouby2010towards}
Yaghouby, F., Ayatollahi, A., Bahramali, R., Yaghouby, M., Alavi, A.H.: Towards
  automatic detection of atrial fibrillation: {A} hybrid computational
  approach.
\newblock Computers in Biology and Medicine \textbf{40}(11), 919--930 (2010)

\bibitem{zabihi2017detection}
Zabihi, M., Rad, A.B., et~al.: Detection of atrial fibrillation in {ECG}
  hand-held devices using a random forest classifier.
\newblock 2017 Computing in Cardiology (CinC) \textbf{44}, 1--4 (2017)

\bibitem{zhao2018spectro}
Zhao, Z., S{\"a}rkk{\"a}, S., Rad, A.B.: Spectro-temporal {ECG} analysis for
  atrial fibrillation detection.
\newblock In: 2018 IEEE 28th International Workshop on Machine Learning for
  Signal Processing (MLSP), pp. 1--6. IEEE (2018)

\bibitem{zihlmann2017convolutional}
Zihlmann, M., Perekrestenko, D., Tschannen, M.: Convolutional recurrent neural
  networks for electrocardiogram classification.
\newblock 2017 Computing in Cardiology (CinC) \textbf{44}, 1--4 (2017)

\bibitem{zoni2014epidemiology}
Zoni-Berisso, M., Lercari, F., Carazza, T., Domenicucci, S.: Epidemiology of
  atrial fibrillation: {E}uropean perspective.
\newblock Clinical Epidemiology \textbf{6}, 213--220 (2014)

\end{thebibliography}

\end{document}